\documentclass[12pt,preprint]{aastex}  %  one-column, single-spaced

\usepackage{natbib,amsmath}
\newcommand{\parcomp}{Parallel~Computing}
\newcommand{\sci}{Science}

\newcommand{\arfm}{Ann.~Rev.\ Fluid~Mech.}

\newcommand{\sone}{S1}  %%  case S1
\newcommand{\stwo}{S2}  %%  case S2
\newcommand{\dtwo}{D2}  %%  case D2
\newcommand{\dthr}{D3}  %%  case D3

\newcommand{\bcr}{\bs\times} %% bold vector cross product operator
\newcommand{\bdel}{\bs\nabla} %% bold del operator
\newcommand{\bdot}{\bs\cdot} %% bold dot operator
\newcommand{\bs}{\boldsymbol} %% for putting things in math bold
\newcommand{\bu}{\bs u} %% bold u (velocity)
\newcommand{\deldot}{\bdel\bdot} %% divergence operator
\newcommand{\pd}{\partial} %% partial derivative symbol
\newcommand{\rhat}{\,\bs{\hat{r}}}  %%  r-hat unit vector
\newcommand{\udotdel}{(\bu\bdot\bdel)} %% u dot del

\shorttitle{Multi-Scale Solar Convection}
\shortauthors{DeRosa, Gilman, \& Toomre}

\begin{document}

\title{Solar Multi-Scale Convection and Rotation Gradients
Studied in Shallow Spherical Shells}

\author{Marc L. DeRosa\altaffilmark{1}} 

\affil{JILA and Department of Astrophysical and Planetary Sciences\\
University of Colorado, Boulder, CO 80309--0440}

\altaffiltext{1}{Current address: Lockheed Martin Solar and Astrophysics
Laboratory, Bldg.~252, Org.~L9--41, 3251 Hanover St., Palo Alto, CA 94304}
\email{derosa@lmsal.com}

\author{Peter A. Gilman} 

\affil{High Altitude Observatory, National Center for Atmospheric
Research\altaffilmark{2} \\ P.O. Box 3000, Boulder, CO 80307--3000}

\altaffiltext{2}{The National Center for Atmospheric Research is sponsored by
the National Science Foundation.}

\and

\author{Juri Toomre}

\affil{JILA and Department of Astrophysical and Planetary Sciences\\ University
of Colorado, Boulder, CO 80309--0440}

\begin{abstract}

The differential rotation of the sun, as deduced from helioseismology,
exhibits a prominent radial shear layer near the top of the convection zone
wherein negative radial gradients of angular velocity are evident in the low-
and mid-latitude regions spanning the outer 5\% of the solar radius.
Supergranulation and related scales of turbulent convection are likely to play
a significant role in the maintenance of such radial gradients, and may
influence dynamics on a global scale in ways that are not yet understood.  To
investigate such dynamics, we have constructed a series of three-dimensional
numerical simulations of turbulent compressible convection within spherical
shells, dealing with shallow domains to make such modeling computationally
tractable.  In all but one case, the lower boundary is forced to rotate
differentially in order to approximate the influence that the differential
rotation established within the bulk of the convection zone might have upon a
near-surface shearing layer.  These simulations are the first models of solar
convection in a spherical geometry that can explicitly resolve both the
largest dynamical scales of the system (of order the solar radius) as well as
smaller-scale convective overturning motions comparable in size to solar
supergranulation (20--40~Mm).  We find that convection within these
simulations spans a large range of horizontal scales, especially near the top
of each domain where convection on supergranular scales is apparent.  The
smaller cells are advected laterally by the the larger scales of convection
within the simulations, which take the form of a connected network of narrow
downflow lanes that horizontally divide the domain into regions measuring
approximately 100--200~Mm across.  We also find that the radial angular
velocity gradient in these models is typically negative, especially in the
low- and mid-latitude regions.  Analyses of the angular momentum transport
indicates that such gradients are maintained by Reynolds stresses associated
with the convection, transporting angular momentum inward to balance the
outward transport achieved by viscous diffusion and large-scale flows in the
meridional plane, a mechanism first proposed by \citet{fou1975} and tested by
\citet{gil1979}.  We suggest that similar mechanisms associated with
smaller-scale convection in the sun may contribute to the maintenance of the
observed radial shear layer located immediately below the solar photosphere.

\end{abstract}

\keywords{convection --- hydrodynamics --- Sun: interior --- Sun: rotation ---
turbulence}

\section{Introduction}

\notetoeditor{I am ok with single-column figures, except where I have
indicated otherwise.  Figures 5 -- 10, 12, 13, 15, and 17 should be in color.}

\begin{figure}
  \plotone{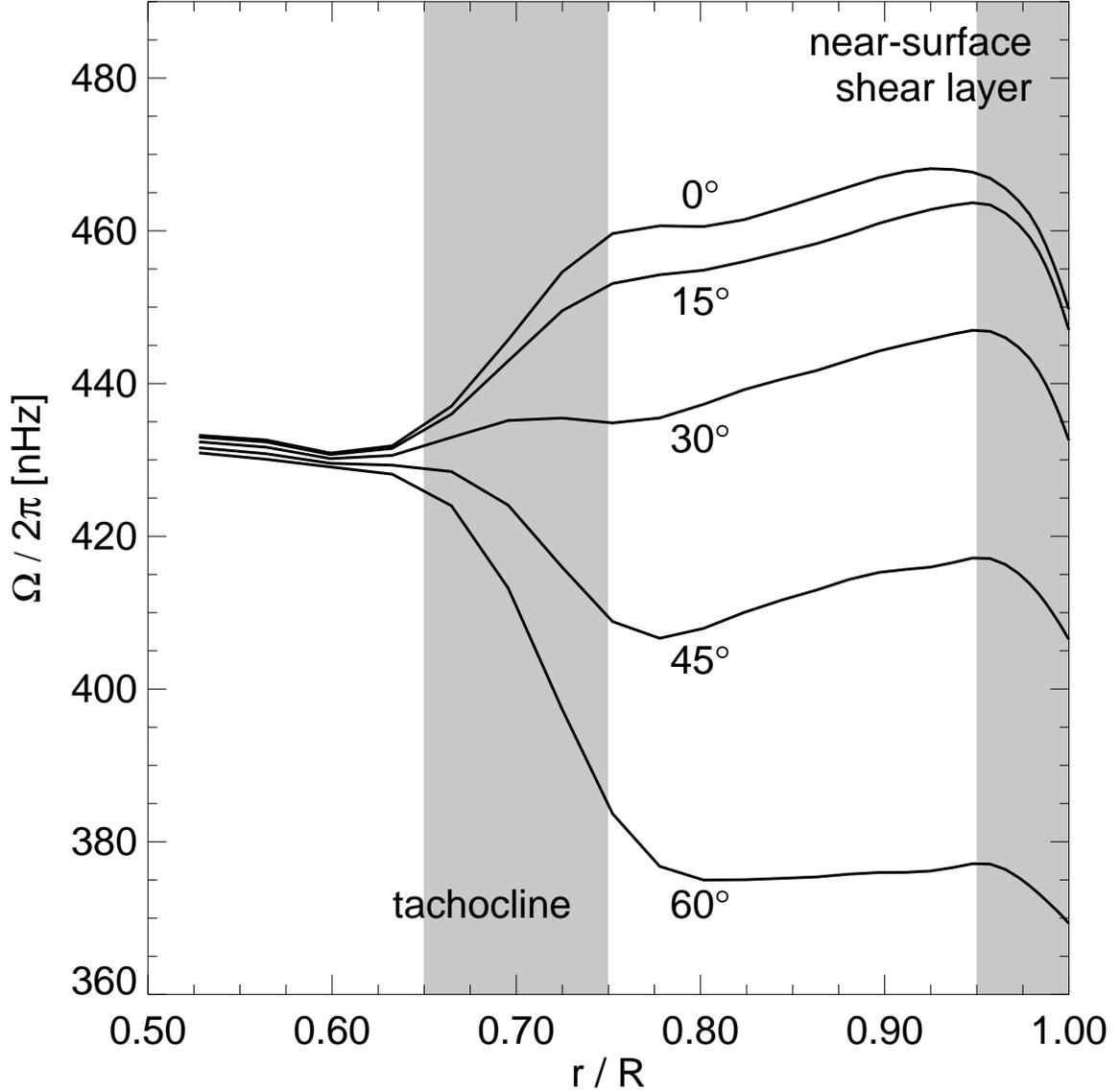}
  \caption{Variation of angular velocity $\Omega/2\pi$ with proportional
  radius $r/R$ at selected latitudes as inferred from helioseismic RLS
  inversions averaged over four years of GONG data (adapted from
  \citealt{how2000a}).  Shear layers ({\sl shaded}), as evidenced by more rapid
  variations of $\Omega$ with radius, are observed near the base of the
  convection zone as well as near the surface, with the latter region
  extending from 0.95~$R$ to 1.00~$R$.  The gradients of $\Omega$ in that
  near-surface shear layer at high latitudes are sensitive to the
  inversion method and data sets used \citep{sch2002}}  \label{fig:gong}
\end{figure}

Helioseismology has revealed that the differential rotation profile observed
at the solar photosphere roughly extends throughout the bulk of the
convection zone \citep{tho1996,sch1998b}.  From about 0.75~$R$ to 0.95~$R$
(where $R$ is the solar radius), the angular velocity $\Omega$ has a small
radial gradient, particularly at mid latitudes, as seen in
Figure~\ref{fig:gong}.  In contrast, regions of strong radial shear are
evident near both the bottom and top of the convection zone (shown shaded in
Fig.~\ref{fig:gong}), and these shear layers are believed to play important
dynamical roles within the solar convection zone.  While the tachocline
region at the base of the convection zone has commanded much recent attention
(as it is likely the seat of the global solar dynamo and the associated
22-year magnetic activity cycle), the dynamics within the near-surface shear
layer, extending from 0.95~$R$ to 1.00~$R$, are also likely to have
additional dynamical consequences that affect the appearance and evolution of
flows and magnetic structures visible at the surface.  Such dynamics are
presently not well understood, but are now becoming accessible to study
through direct numerical simulations that capture the important effects of
sphericity, compressibility, and rotation.

Several questions arise about the near-surface shear layer.  First, what
dynamical mechanisms within the coupling of turbulent convection with
rotation leads to such a boundary layer involving negative radial gradients
of $\Omega$?  In contrast, in the bulk of the convection zone the gradients
are much smaller and positive.  Second, why does this boundary layer have a
depth of about 5\% in solar radius?  Third, does the presence of such a
strong radial shear zone play a significant role in the complex large-scale
meandering flows and reversing meridional circulations that have been shown
by helioseismology to coexist with the intense smaller-scale convection in
the upper reaches of the convection zone?   These three issues motivate our
studies as we seek to resolve both supergranulation and global-scale
responses in our simulations of convection in rotating spherical shells.
Computational constraints have encouraged us to begin by studying thin shells
of such turbulent convection, encompassing at this stage only the upper
portion of the solar convective envelope.  A preliminary account of such
modeling is presented in \citet{der2001}.  We shall here show that the
resulting multi-scale convection is able to redistribute angular momentum so
as to yield radial gradients in $\Omega$ that are largely in accord with the
helioseismic findings.  These first steps are important in defining more
complex simulations to be undertaken within deep convection shells that
capture much of the depth range of the solar convection zone.

The stratification within the near-surface shear layer serves to drive
vigorous motions possessing a wide range of spatial and temporal scales,
visible at the surface as the convective patterns of supergranulation,
mesogranulation, and granulation \citep{spr1990}. Many aspects of such
small-scale but intensely turbulent convection, influenced by radiative
transfer effects and complex equations of state and opacities, have been
studied through three-dimensional simulations within localized planar domains
positioned near the solar surface (e.g.~\citealt{ste1998,ste2000,ste2001}).
The driving in such convection is enhanced by the latent heat released within
the ionization zones of helium and hydrogen that are present in the
near-surface shear layer (e.g.~\citealt{ras1993}).  These small-scale
turbulent convective motions are likely to facilitate the transport of angular
momentum along both radial and latitudinal velocity gradients within the shear
layer, and thus may be able to affect the dynamics within the convection zone
on a more global scale.  In particular, the horizontal extent and overturning
time of supergranular flows suggest that such convection will be at least
weakly influenced by rotational effects, which can yield Reynolds stresses of
significance in transporting angular momentum within the layer.  The coupling
of turbulent compressible convection with rotation has also been studied
extensively in localized $f$-plane domains
(e.g.~\citealt{bru1995,bru1996,bru1998,brum2002,bra1996,cha2001}) using
perfect gases, revealing that the presence of coherent structures associated
with strong downflow plumes and networks play a crucial role in the
redistribution of angular momentum.  Such studies are complemented by a broad
range of other simulations of compressible convection that exhibit intrinsic
asymmetries between upflows and downflows, of complex vorticity structures
that influence the transport of heat, momentum and magnetic fields, and of
rich time-dependence involving a broad range of time scales
(e.g.~\citealt{cat1991,por1994,por2000,sai2000,rob2001,tob2001}).

Velocity features larger than the spatial scale of solar supergranulation are
also in evidence in the near-surface shear layer. Bands of slightly faster
rotation, or torsional oscillations, that gradually propagate toward the
equator as the magnetic activity cycle advances are detected in Doppler
measurements of the surface \citep{lab1982,hat1996,ulr1998}.  They are also
seen in global $f$- and $p$-mode helioseismic studies
(e.g.~\citealt{kos1997,sch1999,how2000b,vor2002}), and are present over at
least the outer 8\% in radius.  Even more complex flows within the
near-surface shear layer, now called solar subsurface weather \citep{too2002},
are revealed by local-domain helioseismic techniques such as ring-diagram
analyses (e.g.~\citealt{hil1988,hab1998,hab2000,bas1999}) and time-distance
methods (e.g.~\citealt{duv1993,gil1997,duv2000,cho2001}).  Mappings of
subsurface flow fields over a range of depths reveal evolving large-scale
horizontal flows that are somewhat reminiscent of jet streams, meridional
circulations that may possess multi-celled structures in one hemisphere and
not in the other, and distinctive flow deflection in the vicinity of active
complexes \citep{hab2002}.  Although the flow speeds in the meridional
circulations are only of order 20~m~s$^{-1}$, they may be quite effective in
redistributing angular momentum in latitude, thereby coupling widely separated
regions within the near-surface shear layer and possibly having a role in its
existence.

Photospheric magnetic field observations reveal structured concentrations that
also possess a wide range of spatial scales, including active regions,
sunspots, pores, and emergent flux elements.  On the smallest observable
scales, concentrations of filamentary magnetic flux elements are found to be
laterally advected by larger-scale surface flow patterns.  Outflows associated
with the convective patterns of supergranulation and granulation in particular
are observed to readily advect such small-scale flux toward intercellular
lanes and concentrate these fields on scales small enough for dissipation to
occur \citep{sch1997,ber1998}.

Figure~\ref{fig:gong} indicates that the radial gradient of angular velocity
$\Omega$ is largely negative at low and mid latitudes within the near-surface
shear layer, such that the rotation rate decreases by about 2--4\% as one
moves outward across the layer.  Such a radial gradient in $\Omega$ may be
interpreted as a tendency for fluid parcels in the convection zone to
partially conserve their angular momentum as they move toward or away from
the axis of rotation.  This idea was originally suggested by \citet{fou1975},
and may explain why larger-scale magnetic tracers at the surface have a
faster rotation rate relative to the photospheric plasma, assuming that these
magnetic features are anchored at a radius below the photosphere where the
rotation rate is faster.  Numerical simulations  of Boussinesq fluids
confined to thin shells \citep{gil1979} showed that angular momentum is
roughly conserved along radial lines, and small-scale convective motions are
able to transport angular momentum inward, thereby maintaining the negative
radial gradient of rotation rate with radius for such an incompressible
fluid.  Whether the same is true for a compressible fluid is one of the
topics addressed in this paper.

The maintenance of the relatively small radial gradients of $\Omega$
throughout the bulk of the convection zone must be a direct consequence of
the interaction of rotation with the turbulent fluid motions that exist
within the solar convection zone.  Recent three-dimensional numerical
simulations of such deep convection within rotating spherical shells
(e.g.~\citealt{mie2000,ell2000,bru2001a,bru2002}) indicate that for a range
of parameter values, solar-like differential rotation profiles can be
established, even with viscous and thermal diffusivities in a regime far
removed from their solar values.  Many of these simulations possess about a
30\% contrast in angular velocity $\Omega$ between the equator and high
latitudes and have small radial gradients of $\Omega$ in the mid-latitude
regions, features that roughly match the helioseismic determinations of the
interior rotation profile within the bulk of the convection zone.  Analyses
of the angular momentum transport within these simulations indicate that the
fast equatorial rotation relative to the higher latitudes is primarily
maintained by a complex interplay between global meridional circulation and
Reynolds stresses achieved within the domains, both of which contribute to
the equatorward transport of angular momentum with latitude.

The radial velocity planforms within the more laminar convection zone
simulations take the form of rotationally aligned banana cell structures,
with the downflowing fluid lanes extending throughout most of the radial
extent of the domain.  As the level of turbulence is increased, these
organized patterns become less prominent, giving way to a network of narrower
downflow lanes that form plume-like structures at the interstices in the
network.  Such plumes tend to possess significant vortical motion and span
the entire domain in radius.  The influence of rotation on these vertical
plumes preferentially tilts these structures such that they are partially
aligned with the axis of rotation, which in turn creates the Reynolds
stresses that facilitate angular momentum transport within the domain.  These
deep-shell simulations of the convection zone typically place the upper
boundary at about 0.96~$R$, and thus do not capture the smaller scales of
convection that exist closer to the surface.  Consequently, the convection in
even the most turbulent of these simulations involves overall pattern scales
of order $20^\circ$--$30^\circ$, or several hundred~Mm, although the sheets
and plumes associated with the downflow network are individually narrower and
more concentrated.

To understand more clearly some of the physical processes occurring within
the near-surface shear layer of the solar convection zone, we have
constructed numerical simulations of compressible fluids within thin
spherical shells that extend up to 0.98~$R$, encompassing solely the
near-surface shear layer region.  Continual advances in supercomputing
technology now permit three-dimensional compressible fluid simulations that
explicitly resolve spatial and temporal scales spanning several orders of
magnitude.  As a result, we are able for the first time to employ direct
numerical simulations to investigate the effects of supergranular-sized
convection on the more global dynamics within the near-surface shear layer of
the sun.

In formulating our simulations, we have adopted the viewpoint that the
latitudinal variation of the angular velocity in the sun, with the equatorial
regions rotating more rapidly than the poles as in Figure~\ref{fig:gong}, is
established and maintained within the bulk of the convection zone somewhere
below the lower boundary of our thin shell models.  We have thus imposed a
solar-like differential rotation profile as a no-slip lower boundary in three
of the four simulations presented here, in order to capture some of the
dynamical effects related to such an angular velocity structure.  In so
doing, we are implicitly assuming that the global differential rotation
profile is not substantially affected by the convection within our thin
shells, even though in the sun such shearing layers could have subtle effects
on these dynamics.

Our primary focus in this paper will be to investigate the angular momentum
transport achieved by multi-scale convection, involving both global and
supergranular scales, within shearing layers analogous to the near-surface
shear layer of the sun.  We shall consider radial stratifications that
resemble ones deduced from stellar structure models over the depth range
being studied, though the physics of the gases is highly simplified.  After
briefly discussing in \S\ref{sec:ashcode} the governing equations and
numerical approach used in solving them, we review the parameters used to
initialize our thin-shell simulations in \S\ref{sec:thinshellsetup}.  We next
examine in \S\ref{sec:modelflows} the multi-scale convective velocity
patterns of the mature solutions, and discuss the meridional circulation,
time-evolution and angular momentum balance achieved within the thin shell
domains.  Lastly, we discuss the connection between these simulations and the
near-surface shear layer of the solar convection zone, and present possible
directions for future research in \S\ref{sec:conc}.

\section{Governing Equations and Numerical Approach} \label{sec:ashcode}

The convection simulations described here are carried out using the anelastic
spherical harmonic (ASH) computer code \citep{clu1999}, which solves the
anelastic equations of hydrodynamics describing a compressible fluid confined
to a spherical shell heated from below.  The fluid motions are calculated with
respect to a rotating frame.  The complex structures and intricate behavior of
the resulting convection require high spatial resolution, and the flows must
be studied over extended periods of time for statistical equilibrium to be
achieved.  As a result, the ASH code is designed to run efficiently on
massively parallel architectures such as the Cray~T3E, SGI~Origin2000, and
IBM~SP-3 machines (e.g.~\citealt{mie2000,ell2000}).  The current
multi-processor version follows the numerical approach first implemented by
\citet{gla1984}.

The ASH code employs a pseudo-spectral approach, where all fluid velocities
and state variables are projected onto orthogonal basis functions in each of
the three spatial dimensions.  The radial structure of the solution variables
is represented by an expansion based on Chebyshev polynomials characterized by
radial order $n$, while functions of latitude and longitude are expanded over
spherical harmonic basis functions $Y_\ell^m$ characterized by angular degree
$\ell$ and azimuthal order $m$.  This discretization scheme ensures that the
horizontal resolution is approximately uniform everywhere on the spherical
domain when all $(\ell,m)$-pairs for a given maximum degree $\ell_{\max}$ are
retained in the modal expansion (such as the triangular truncation used here).
Conversely, the simplest finite-difference scheme, where computational
gridpoints are uniformly distributed along lines of latitude and longitude,
suffers from the problem that the spatial resolution varies with latitude such
that the gridpoints are more closely spaced near the poles compared to
equatorial regions (colloquially known as the {\sl pole problem}).  In
addition, spectral methods also provide increased computational accuracy for a
given grid size when compared to finite-difference representations.

The ASH code solves an approximate form of the Navier-Stokes equations known
as the anelastic equations.  The anelastic approximation \citep{gou1969}
allows us to handle the effects of compressibility while filtering out
acoustic perturbations that would otherwise severely limit the computational
time step.  This approximation is valid when the convective fluid velocities
are subsonic, which in turn requires that the stratification of the fluid be
only slightly superadiabatic.  Such acoustic filtering is achieved by
insisting that the time derivative of density vanishes in the continuity
equation, or equivalently that the divergence of the momentum be zero, or that
the momentum vector be solenoidal.  This approximation is effectively
equivalent to allowing pressure disturbances to equilibrate instantaneously,
forcing the system to evolve on convective rather than sound-speed time
scales.  It is therefore implicitly assumed that sound waves do not play a
significant role in the dynamical evolution of the system, which is in
agreement with the expectation that the coupling of convection,
stratification, and rotation are the major dynamical influences within the
convection zone.

The convective motions are computed relative to a spherically symmetric mean
state having a density~$\bar{\rho}$, temperature~$\bar{T}$,
pressure~$\bar{p}$, and specific entropy~$\bar{s}$, much as discussed in
\citet{mie2000,bru2002}.  These mean quantities initially satisfy the equation
of hydrostatic equilibrium,
\begin{equation}
  \frac{d\bar{p}}{dr} = -\bar{\rho} g, \label{eq:hydrostatic}
\end{equation}
where $g$ is the acceleration due to gravity, and a perfect-gas equation of
state
\begin{equation}
  \bar{p} = \frac{\gamma-1}{\gamma}\, c_p \bar{\rho} \bar{T},
  \label{eq:meaneos1}
\end{equation}
from which the mean specific entropy $\bar{s}$ is defined (to within an
arbitrary constant) by
\begin{equation}
  \frac{d\bar{s}}{dr} = c_p \left( \frac{1}{\gamma \bar{p}}\,
  \frac{d\bar{p}}{dr} - \frac{1}{\bar{\rho}}\, \frac{d\bar{\rho}}{dr}
  \right), \label{eq:meaneos2}
\end{equation}
where the specific heats at constant pressure and volume are represented by
$c_p$ and $c_v$ respectively, with their ratio $\gamma = c_p/c_v$.  The mean
thermodynamic state is allowed to evolve in time.

The anelastic approximation involves neglecting the time-derivative of density
in the mass continuity equation (as described above), such that
\begin{equation}
  \deldot (\bar{\rho} \bu) =0,
\end{equation}
and expanding the Navier-Stokes equations around the spherically symmetric
mean state.  Perturbations to the state variables are denoted by $\rho$, $T$,
$p$, and~$s$.  The ASH code solves the equations describing the evolution of
the fluid velocity,
\begin{equation}
  \bar{\rho}\frac{\pd\bu}{\pd t} = -\bar{\rho} \udotdel \bu + 2\bar{\rho}
  (\bu\bcr\bs\Omega_0) -\bdel p -\rho g \rhat +\deldot \bs{\mathcal D},
  \label{eq:evol1}
\end{equation}
and of the specific entropy,
\begin{equation}
  \bar{\rho} \bar{T} \frac{\pd s}{\pd t} = - \bar{\rho} \bar{T} \udotdel
  (\bar{s} + s) + \deldot \bs q + \Phi, \label{eq:evol2}
\end{equation}
where the viscous stress tensor $\bs{\mathcal{D}}$ and viscous heating term
$\Phi$ are defined
\begin{equation}
  \mathcal{D}_{ij} = 2 \bar{\rho} \nu \left[ e_{ij} - \frac{1}{3} (\deldot\bu)
  \delta_{ij} \right]
\end{equation}
and
\begin{equation}
  \Phi = 2 \bar{\rho} \nu \left[ e_{ij} e_{ij} - \frac{1}{3} (\deldot\bu)^2
  \right],
\end{equation}
with $e_{ij}$ denoting the strain rate tensor.  The diffusive heat flux $\bs
q$ is defined
\begin{equation}
  \bs q = \kappa_s \bar{\rho} \bar{T} \bdel (\bar{s} +s) + \kappa_r c_p
  \bar{\rho} \bdel (\bar{T} + T).
\end{equation}
Furthermore, these equations are subject to the linearized equation of state,
\begin{equation}
  \frac{\rho}{\bar{\rho}} = \frac{p}{\bar{p}} - \frac{T}{\bar{T}} =
  \frac{p}{\gamma \bar{p}} - \frac{s}{c_p}.
\end{equation}

As with the temporal scales of motion, fully resolving all spatial scales of
motion in a numerical simulation of the solar convection zone is infeasible at
this time, as the dynamically active scales range from $10^2$~Mm (of order the
depth of the zone) to $10^{-4}$~Mm (typical dissipation scale), thereby
encompassing a factor of at least $10^6$ in scale.  Because current
simulations can cope with a range of only about $10^3$ in each of the three
physical dimensions, the ASH code adopts the common approach of parameterizing
the transport properties of sub-grid scale (SGS) turbulent eddies and
resolving only the largest scales of convection, thus becoming a large eddy
simulation (LES).

All LES-SGS simulations require a prescription for representing the effects
of SGS convective motions not explicitly resolved in the model.  Such a
scheme may incorporate characteristics of the resolved flows into their
functional forms (see the reviews by \citealt{can1996,les1997,can1998}), or
may simply enhance the viscous and thermal diffusivities relative to their
molecular values. We have adopted the latter approach for simplicity, yet
recognize that this aspect requires considerable attention in the future. The
main drawback of this scheme is that the enhanced diffusion draws energy from
larger resolved scales of motion which should be unaffected by such
dissipative effects.  In one alternative approach, known as hyperviscosity,
one allows the enhanced eddy diffusivities to act on fourth- (or higher-)
order derivatives of the velocity field, thereby confining the diffusive
effects more toward the smaller end of the energy spectrum. Another class of
SGS models involves adding extra stress terms to the equations of motion.
Evolution equations for these additional contributions can then be
constructed once functional forms for the correlations between second-order
variables are specified using some kind of a closure hypothesis. As is true
of all LES-SGS studies, one hopes that the specific form by which the SGS
motions are parameterized has a relatively small effect on the global
dynamics of the system, but this is a property that cannot be readily
verified at this stage.

Time-stepping in the ASH code is performed using an implicit second-order
Crank-Nicholson procedure for the linear terms and a fully explicit
second-order Adams-Bashforth procedure for the nonlinear terms, with the
exception of the linear Coriolis forcing terms in the momentum equation which
are updated explicitly.  Because the explicit time-stepping procedure cannot
be performed in the spectral domain, such a pseudo-spectral scheme
necessitates conversions between the physical and spectral representations
during each time step when switching between solving the implicit and explicit
terms in the evolution equations.  However, the benefits gained by solving the
equations in the spectral domain outweigh the added computational time spent
in performing the transformations between the physical and spectral domains.

\section{Model Formulation}  \label{sec:thinshellsetup}

\begin{deluxetable}{lcccc}

%\tabletypesize{\scriptsize}

\tablecaption{A summary of the parameters of the thin-shell convection
simulations.  The nondimensional fluid parameters $T_a$, $R_e$, and $R_a$ are
defined as in \citet{mie2000}, and are evaluated at the middle of the layer.
The quantities $N_r$,~$N_\theta$, and~$N_\phi$ denote the number of
computational gridpoints in physical space across in $r$,~$\theta$, and~$\phi$
respectively.}
\label{ta:simsum}

\tablehead{\colhead{Parameter} & \colhead{Case~\sone} & \colhead{Case~\stwo} &
\colhead{Case~\dtwo} & \colhead{Case~\dthr}}

\startdata

Radial Extent & 0.94--0.98~$R$ & 0.94--0.98~$R$ &
0.90--0.98~$R$ & 0.90--0.98~$R$ \\

Shell Thickness [Mm] & 28 & 28 & 56 & 56 \\

Angular~Velocity $\Omega_0/2\pi$ [nHz] & 410 & 410 & 410 & 410 \\

Rotation~Period $2\pi/\Omega_0$ [days] & 28.2 & 28.2 & 28.2 & 28.2 \\

Shell Density Contrast & 7.5 & 7.5 & 18 & 18 \\

$\nu_{\text{top}}$ [cm$^2$~s$^{-1}$] & $1\times10^{12}$ & $1\times10^{12}$ &
$1\times10^{12}$ & $1\times10^{12}$ \\

$\kappa_{\text{top}}$ [cm$^2$~s$^{-1}$] & $1\times10^{12}$ & $1\times10^{12}$
& $1\times10^{12}$ & $1\times10^{12}$ \\

Prandtl Number $P_r$ & 1 & 1 & 1 & 1 \\

Taylor Number $T_a$ & $5.4\times10^3$ & $5.4\times10^3$ & $2.1\times10^5$ &
$2.1\times10^5$ \\

Reynolds Number $R_e$ & $1.4\times10^2$ & $1.4\times10^2$ & $2.2\times10^2$ &
$2.2\times10^2$ \\

Rayleigh Number $R_a$ & $1.9\times10^4$ & $1.9\times10^4$ & $5.4\times10^5$ &
$5.4\times10^5$ \\

Supercriticality $R_a/R_{a,0}$ & 40 & 100 & 500 & 500 \\

Averaging Interval [days] & 140 & 140 & 36 & 10 \\

$N_r\times N_\theta\times N_\phi$ & 64$\times$512$\times$1024 &
64$\times$512$\times$1024 & 128$\times$512$\times$1024 &
128$\times$512$\times$1024 \\

$\ell_{\max}$ & 340 & 340 & 340 & 340 \\

Angular Periodicity & four-fold & four-fold & four-fold & none \\

Lower Boundary Rotation & uniform & differential & differential & differential

\enddata

\end{deluxetable}

\subsection{Initializing the Spherically Symmetric Mean State}

\begin{figure}
  \epsscale{0.7} \plotone{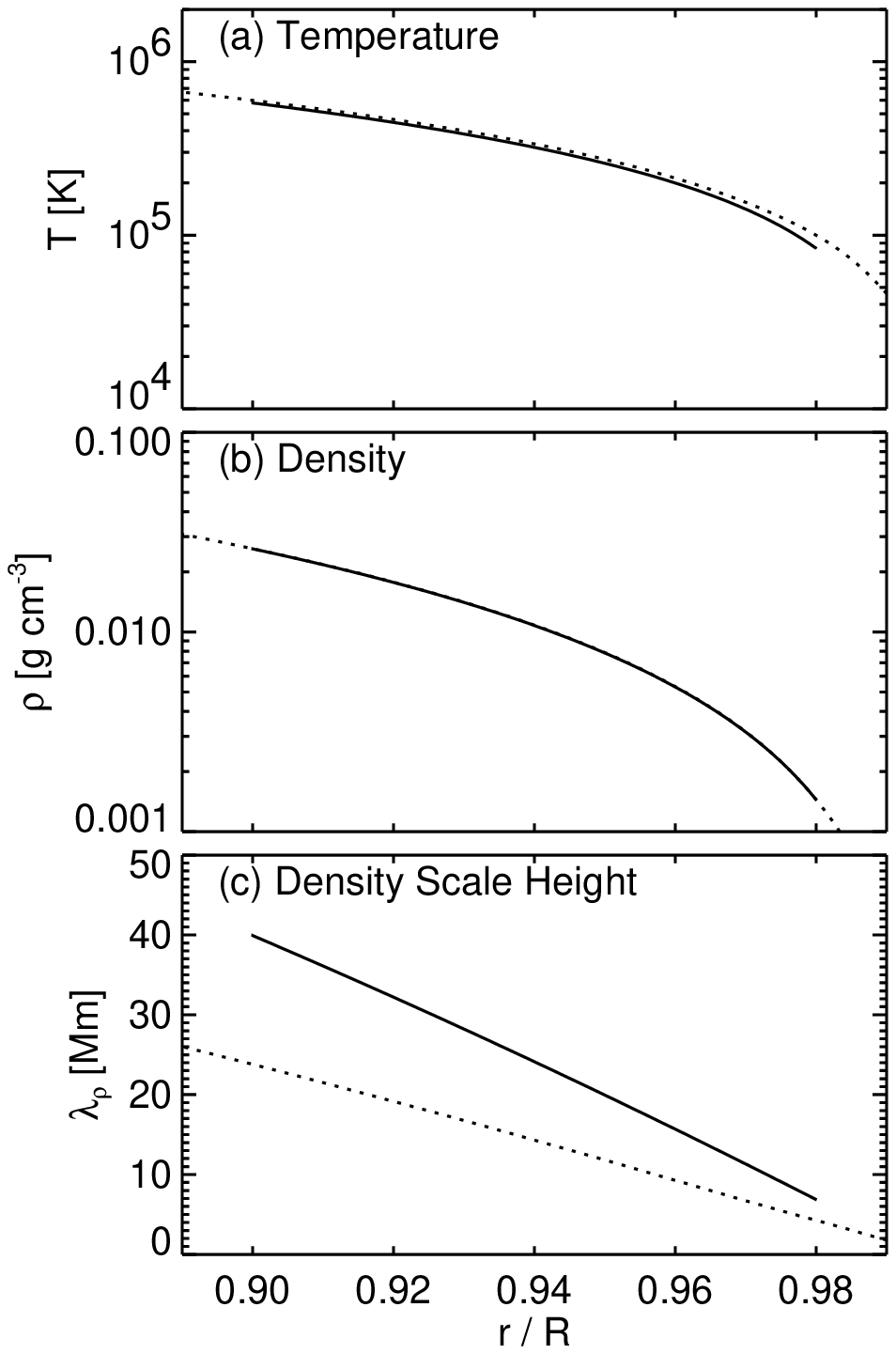}
  \caption{The radial profiles of ($a$) temperature~$\bar{T}$, ($b$)
  density~$\bar{\rho}$, and ($c$) density scale height~$\lambda_\rho$ as a
  function of proportional radius $r/R$ used upon initialization of the
  simulations ({\sl solid lines}), compared with the corresponding values
  taken from a one-dimensional solar structure model ({\sl dotted lines}).}
  \label{fig:refstate}
\end{figure}

We have constructed two shallow-shell simulations (Cases~\sone{} and~\stwo)
that span a radial extent of 0.94--0.98~$R$, equivalent to a shell thickness
of 28~Mm.  Two additional simulations (Cases~\dtwo{} and~\dthr) span
0.90--0.98~$R$ or 56~Mm, and are twice as thick as the two shallow-shell
simulations.  In Cases~\stwo,~\dtwo, and~\dthr, a solar-like differential
rotation profile is applied as a no-slip lower boundary.  For comparison
purposes, the lower boundary of Case~\sone{} is maintained at a uniform
rotation rate equal to the angular velocity $\Omega_0$ of the computational
frame, with all other attributes equivalent to Case~\stwo. In each of the four
cases, the upper boundary is stress-free and both the lower and upper
boundaries are impenetrable.  The thermal driving in each case is accomplished
by setting the heat flux at the lower boundary equal to the solar value, while
the upper boundary is held at a uniform and constant entropy.  The parameters
of the four thin-shell simulations are summarized in Table~\ref{ta:simsum}.

As discussed in the previous section, the anelastic equations of motion are
advanced in time by the ASH code using a pseudo-spectral approach, wherein
functions of colatitude~$\theta$ and longitude~$\phi$ are expanded over
spherical harmonic functions characterized by angular degree $\ell$ and
azimuthal order $m$.  Functions of radius are projected onto Chebyshev
polynomials characterized by radial order $n$.  The four simulations presented
here are calculated using spherical harmonic functions with a maximum angular
degree $\ell_{\max}=340$, so that horizontal scales as small as about 10~Mm
can be explicitly resolved.  The highest order Chebyshev polynomial used is
$n=64$ in Cases~\sone{} and \stwo, and $n=128$ in Cases~\dtwo{} and \dthr.
Since we expect the resulting convection to have a limited longitudinal scale
and seek computational economy, we impose a four-fold azimuthal symmetry by
keeping every fourth $m$ value in the spherical harmonic expansions for
Cases~\sone,~\stwo, and~\dtwo.  Such an imposed symmetry is primarily
noticeable only in the high-latitude regions, where the convergence of
meridional lines near the poles limits the longitudinal scale of the
convective structures to sizes smaller than are present at lower latitudes.
In addition, flows across the pole are not permitted.  For comparison,
Case~\dthr{} was computed without such an angular periodicity, but is
otherwise equivalent to Case~\dtwo.

During initialization, the radial profiles of the mean
density~$\bar{\rho}(r)$, temperature~$\bar{T}(r)$, and pressure~$\bar{p}(r)$
are determined by jointly solving the
equations~(\ref{eq:hydrostatic})--(\ref{eq:meaneos2}), given radial profiles
of the gravitational acceleration $g(r)$ and specific entropy gradient
$d\bar{s}/dr$ throughout the domain. We specify the initial entropy gradient
to have a slightly superadiabatic value (e.g.~$-d\bar{s}/dr=10^{-7}$), while
the function $g(r)$ is taken from the one-dimensional solar model of
\citet{bru1999}.  The initial values of~$\bar{\rho}(r)$ and~$\bar{T}(r)$
obtained in this way are shown in Figure~\ref{fig:refstate} to compare
favorably with the one-dimensional structure model, with the slight
discrepancy in density resulting in a greater density scale height
$\lambda_\rho$ in our simulations than in the structure model.

With our current angular resolution ($\ell_{\max}=340$), we are able to
accommodate convection possessing horizontal size scales as small as a few
angular degrees.  Consequently, we place the upper boundary of each simulation
at 0.98~$R$ where the imposed solar-like stratification should drive
convective structures similar in size to solar supergranulation.  The even
greater degree of stratification present above this radius is likely to drive
modes of convection that are too small in physical size to be resolved with
our current grid size.  In addition, as the convection becomes less efficient
closer to the surface, typical convective velocities become a significant
fraction of the speed of sound thereby making the anelastic approximation
inappropriate.  Furthermore, the ideal gas equation of state and the diffusive
treatment of radiative transfer currently used in these calculations would not
capture the effects resulting from the ionization of hydrogen.  Consequently,
the upper boundary of each simulation is located at 0.98~$R$.

The ASH code is a LES-SGS simulation and thus requires a prescription to
account for the transport of energy and momentum by turbulent motions not
explicitly resolved in the simulations.  We have adopted the simplest approach
of enhancing the molecular values of the viscous and thermal diffusivities,
while recognizing that such an SGS treatment is unlikely to capture all
dynamical effects of small-scale turbulence.  The eddy diffusivities $\nu(r)$
and $\kappa_s(r)$ used in these simulations are chosen to vary inversely as
the square root of the mean density profile, so that $\beta=\frac{1}{2}$ in
\begin{equation}
  \nu=\nu_{\text{top}} \left( \frac{\bar{\rho}}{\bar{\rho}_{\text{top}}}
  \right)^{-\beta} \qquad\text{and}\qquad \kappa_s=\kappa_{\text{top}} \left(
  \frac{\bar{\rho}}{\bar{\rho}_{\text{top}}}
  \right)^{-\beta}. \label{eq:diffscaling}
\end{equation}
This particular value of $\beta$ was chosen to allow some variation of the
dissipation scale with the density scale height $\lambda_\rho$, but at the
same time to prevent numerical instabilities near the bottom of the domain
where the dissipation length scale is smallest.  The free parameters
$\nu_{\text{top}}$ and $\kappa_{\text{top}}$ in
equation~(\ref{eq:diffscaling}) are the viscous and thermal diffusivities at
the upper boundary, which we have set to
$\nu_{\text{top}}=\kappa_{\text{top}}=1\times10^{12}$~cm$^2$~s$^{-1}$ in each
case.  Consequently, the Prandtl number $P_r = \nu/\kappa_s =
\nu_{\text{top}}/\kappa_{\text{top}}$ throughout each domain is unity.  We
note that as computing technology becomes more advanced, the parameterization
scheme used to account for SGS transport effects in simulations of highly
turbulent fluids should become less of an issue because such global
simulations will be able to explicitly resolve the convective motions and
associated energy transport at smaller dynamical scales.

In order to prevent the formation of thin diffusive thermal layers having a
radial thickness below the current radial resolution, we have introduced an
unresolved enthalpy flux to the mean state, much as in \citet{mie2000}.
Without such enhancement, steep entropy gradients would otherwise be required
to carry the imposed heat flux, owing to the small radial fluid velocities
(and thus the convective transport of heat) in the vicinity of the
impenetrable boundaries.

The characteristic size scales of the convective structures that appear near
the top of these thin-shell simulations are somewhat sensitive to the degree
of driving they experience, which is in part determined by the functional form
of the unresolved enthalpy flux.  The form of this quantity essentially
determines what fraction of the total energy flux must be transported via
convection near the top of the domain, which in turn feeds back on the entropy
gradient and affects the convection.  We find that adjusting either the
diffusivities or the superadiabaticity within the domain (that is, altering
the Rayleigh number of the convection) will alter the appearance of the
convection, especially near the upper boundary.  We have chosen a somewhat
gentle functional form for this enthalpy flux, having an $e$-folding depth
away from each boundary of about 0.01~$R$ for both the shallow and deep shell
cases.

Finally, the radiative diffusivity $\kappa_r(r)$ throughout each domain is
taken from the structure model used above; however, we note that the radiative
heat flux is several orders of magnitude smaller in our simulations than
convective heat flux throughout the bulk of the domain (as it is in the sun).

\subsection{Approach to Thermal Equilibrium}

\begin{figure}
  \plotone{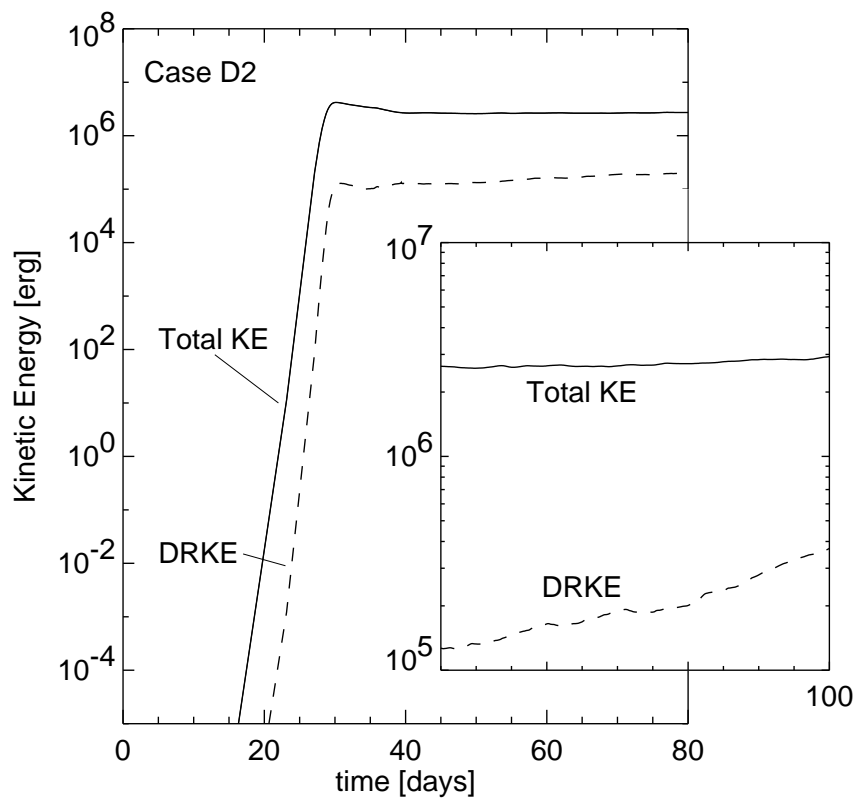}
  \caption{The domain-averaged total kinetic energy ({\sl solid lines}) as a
  function of time for Case~\dtwo.  The ramp-up time of about 30~days from
  seed entropy perturbations is typical of the simulations considered here.
  For comparison, the ramp-up for the kinetic energy ({\sl dashed lines})
  contained in the longitudinally averaged differential rotation (DRKE) is
  also shown.  The inset gives an expanded view of the later evolution, and
  illustrates the continuing growth of both quantities following the initial
  ramp-up phase.  Both quantities are computed relative to the uniform
  rotation of the computational frame.}  \label{fig:approach}
\end{figure}

\begin{figure*}
  \epsscale{0.9} \plotone{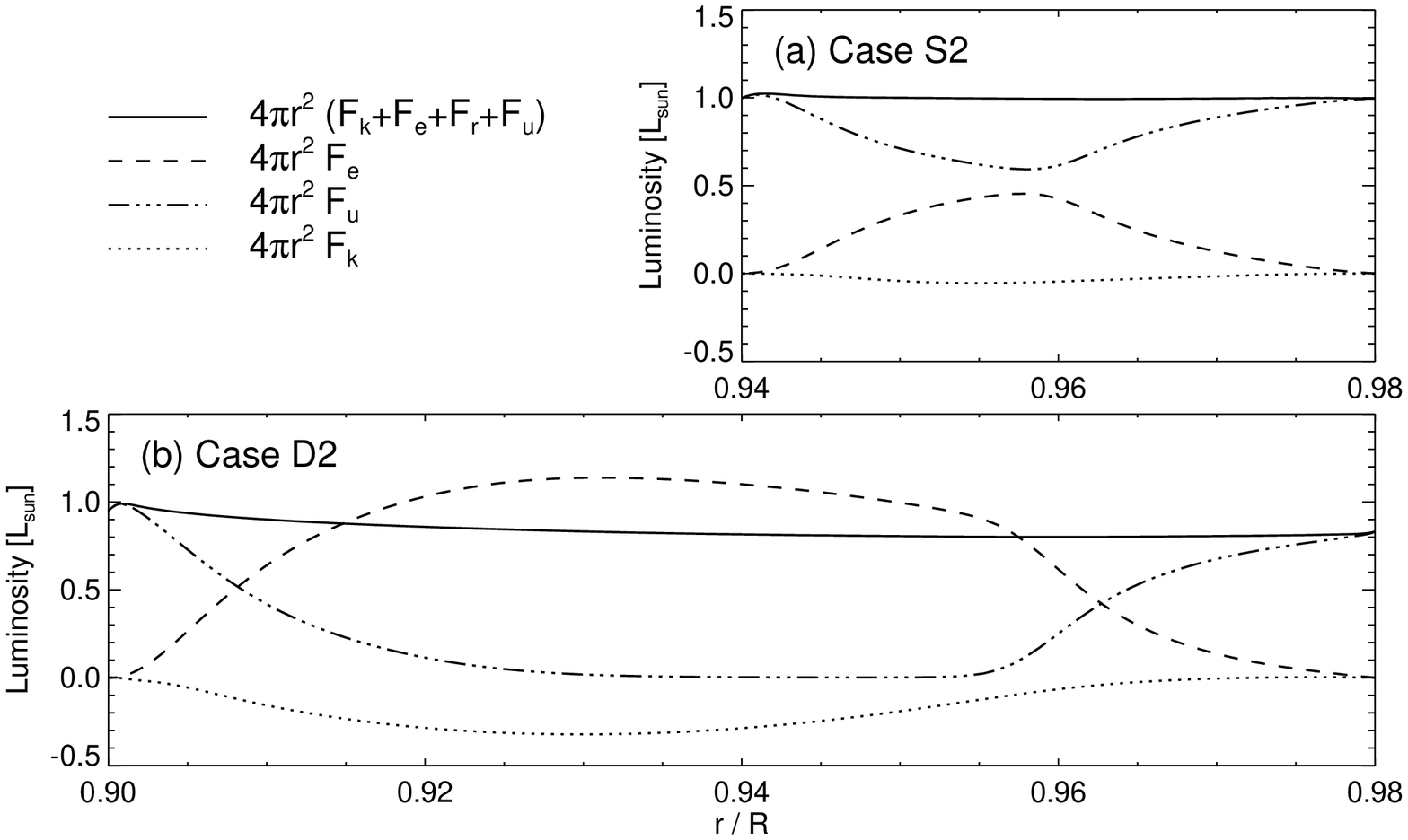}
  \notetoeditor{Figure 4 should be a full-width double-column figure.}
  \caption{The time-averaged spherically symmetric energy balance within
  Cases~\stwo{} and~\dtwo{} as a function of radius, showing the percentage of
  $L_\odot$ carried by kinetic energy ($4\pi r^2\,F_k$), enthalpy ($4\pi
  r^2\,F_e$), and by unresolved eddies ($4\pi r^2\,F_u$).  The radiative
  luminosity ($4\pi r^2\,F_r$) is negligible in both cases and is not shown.}
  \label{fig:fluxbal}
\end{figure*}

Once the spherically symmetric mean state has been arranged, small seed
perturbations are introduced into the fluctuating entropy field $s$, and the
simulations are advanced in time using the evolution
equations~(\ref{eq:evol1}) and~(\ref{eq:evol2}).  The seed entropy
perturbations are soon reflected in the fluctuating density field $\rho$,
whose variations quickly provide the unstable density imbalance that buoyantly
accelerates the fluid from rest.  After an initial period of adjustment during
which the convective kinetic energy ramps up (typically about 30~days, as
shown in Fig.~\ref{fig:approach} for Case~\dtwo), an approximate thermal
equilibrium is reached.  The mean thermodynamic state will evolve in time, as
reflected in the $\ell=0$ component of the thermodynamic variables.

In total thermal equilibrium, the outward energy transport in these
simulations must be achieved by a balance of radiative, kinetic, enthalpy, and
eddy diffusive fluxes, as represented by the following:
\begin{align}
  F_k &= u_r \left( \frac{\bar{\rho} u^2}{2} \right), \\
  F_e &= u_r \bar{\rho} c_p (T- \left< T \right>), \\
  F_r &= -\kappa_r \bar{\rho} c_p \frac{\pd (\bar{T}+T)}{\pd r}, \\
  F_u &= -\kappa_s \bar{\rho} \bar{T} \frac{\pd (\bar{s}+s)}{\pd r},
\end{align}
where the kinetic, enthalpy, radiative, and unresolved eddy fluxes are denoted
by~$F_k$,~$F_e$, $F_r$, and~$F_u$ respectively.  The quantity $T-\left< T
\right>$ appearing in the definition of $F_e$ is the temperature excess
relative to the mean (spherically symmetric) value of the temperature field at
each radial level.  In a true steady state, the total energy transport at each
radius within the domain must equal the imposed energy influx at the lower
boundary:
\begin{equation}
  4\pi r^2 (F_k + F_e + F_r + F_u) = L_\odot,
\end{equation}
where $L_\odot$ is the solar luminosity.

Figure~\ref{fig:fluxbal} shows the time-averaged energy transport within
Cases~\stwo{} and~\dtwo{} as a function of radius at a late stage in the
simulations.  In both cases, the enhanced energy transport by the unresolved
eddies near the boundaries reflects the additional enthalpy flux applied to
the mean state.  Figure~\ref{fig:fluxbal} also shows that Case~\dtwo{} is not
yet in total thermal equilibrium, as the energy output at the upper boundary
is approximately 80\% of the input applied to the lower boundary.  The radial
energy transport within Case~\sone{} is qualitatively similar to Case~\stwo,
and likewise for Cases~\dtwo{} and~\dthr.

\section{Characteristics of Resulting Flows}  \label{sec:modelflows}

\subsection{Multi-Scale Convection}  \label{sec:multiscale}

\begin{figure*}
  \plotone{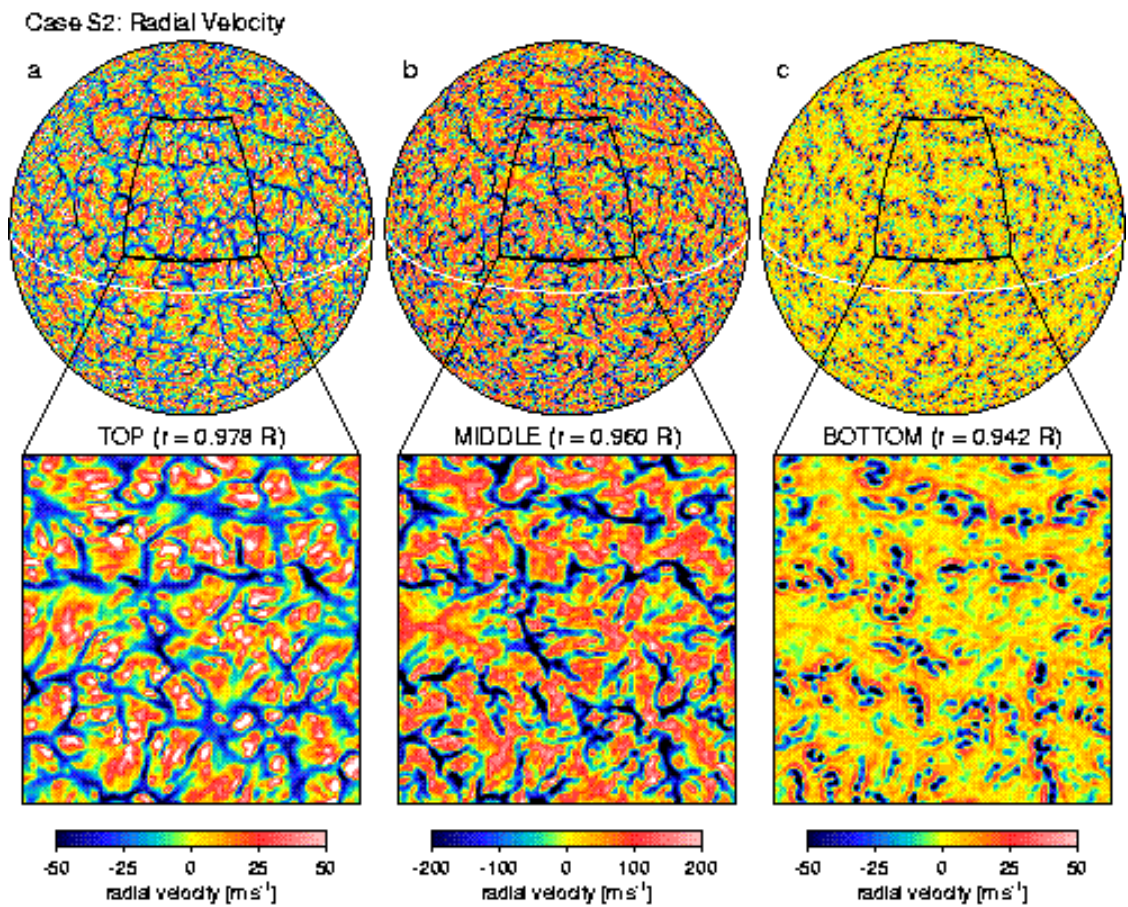}
  \notetoeditor{Figure 5 should be a full-width double-column figure.}
  \caption{Instantaneous snapshots of radial velocity for Case~\stwo{} near
  the ($a$) top, ($b$) middle, and ($c$) bottom of the domain.  Positive
  radial velocities (orange-red colors) denote upflows and negative radial
  velocities (green-blue colors) denote downflows.  Each image in the top row
  is an orthographic projection of the velocity field, with the north pole
  tilted $20^\circ$ toward the observer and the equator indicated by a white
  line.  Each enlarged image in the bottom row shows a rectangular
  (latitude-longitude) projection of a $45^\circ$-square portion of the
  corresponding velocity field in the top row.  The four-fold azimuthal
  periodicity is most noticeable near the north pole.}  \label{fig:shsl_s2_ur}
\end{figure*}

\begin{figure*}
  \plotone{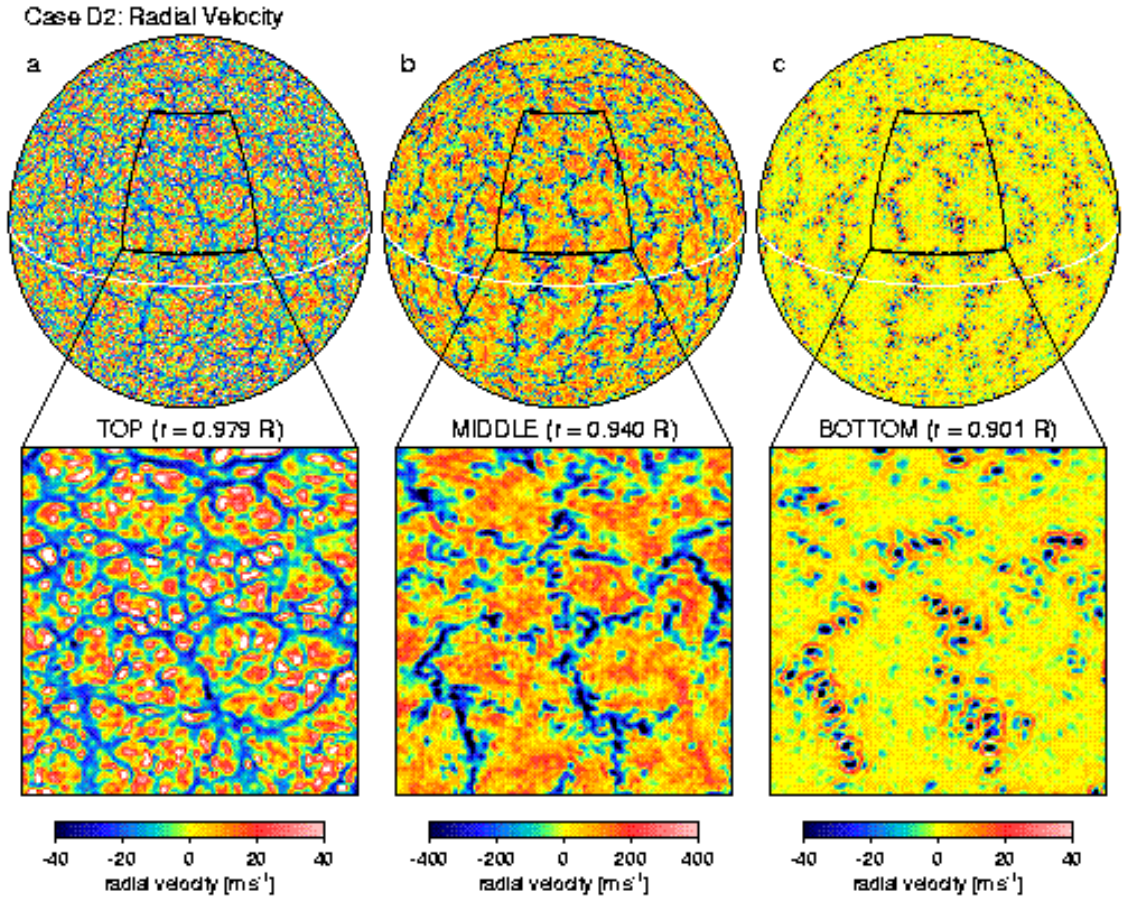}
  \notetoeditor{Figure 6 should be a full-width double-column figure.}
  \caption{Same as in Fig.~\ref{fig:shsl_s2_ur} except for Case~\dtwo, showing
  the radial velocity structure sampled near the ($a$) top, ($b$) middle, and
  ($c$) bottom of the deep-shell simulation.}  \label{fig:shsl_d2_ur}
\end{figure*}

\begin{figure*}
  \plotone{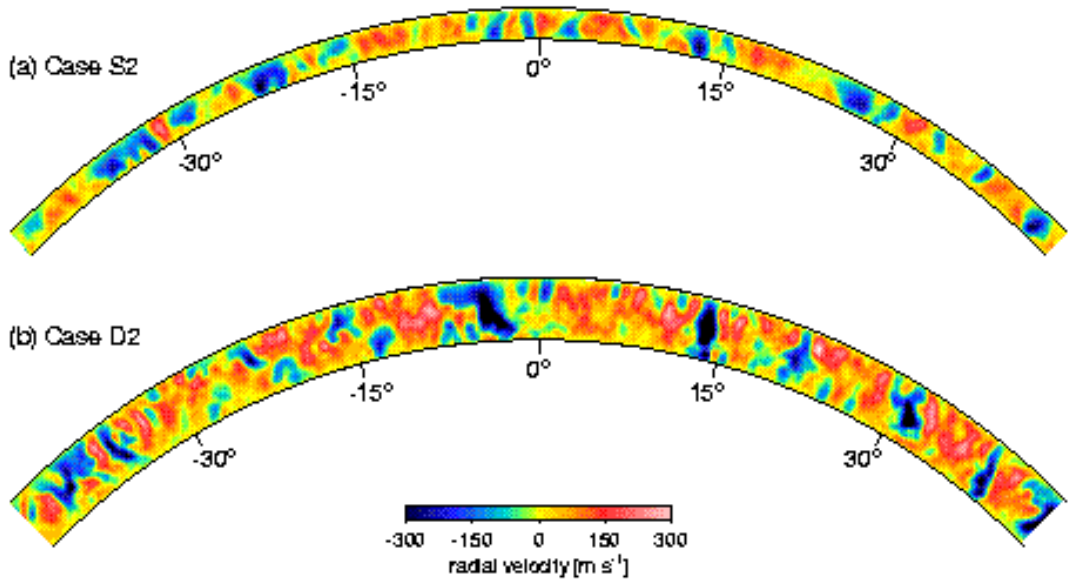}
  \notetoeditor{Figure 7 should be a full-width double-column figure.}
  \caption{Images of the instantaneous radial velocity for Cases~\stwo{}
  and~\dtwo{} as a function of latitude and radius in a cut of fixed
  longitude, showing the vertical structure of the pattern of upflows and
  downflows within each domain.  The fastest small-scale upflows are
  particularly visible near the top of Case~\dtwo.}  \label{fig:posl_ur}
\end{figure*}

\begin{figure*}
  \plotone{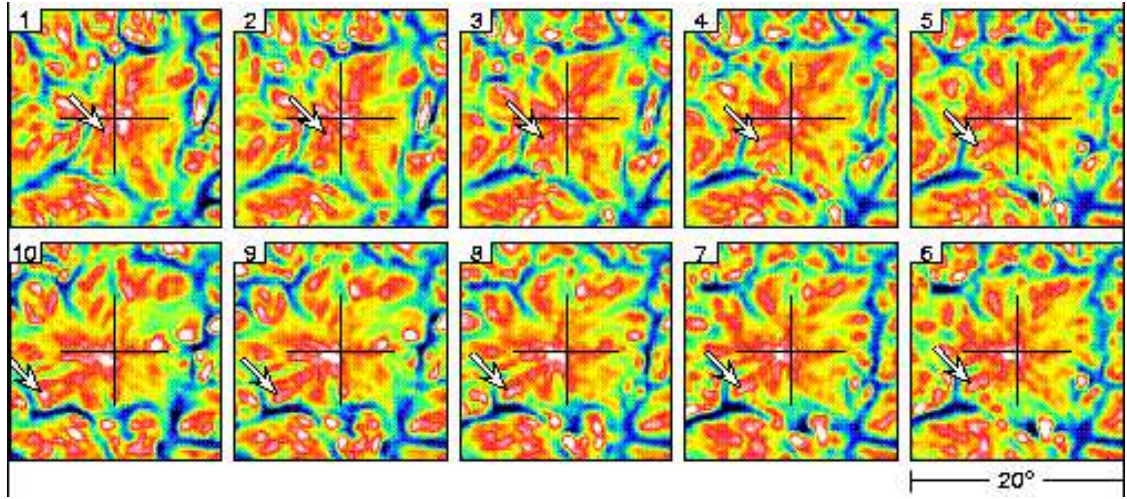}
  \notetoeditor{Figure 8 should be a full-width double-column figure.}
  \caption{A time series of radial velocity images showing a $20^\circ$-square
  region of Case~\stwo{} near the upper boundary in which the lateral
  advection of small-scale features by larger-scale horizontal motions is
  illustrated.  The arrow points to one such small-scale upflow which is
  advected away from the center of the broader upflow (indicated by the dark
  cross).  The time index of each image is indicated in the upper-left corner,
  and the cadence is about 1.3~days between images.}
  \label{fig:evol_shsl_s2_ur}
\end{figure*}

\begin{figure}
  \epsscale{0.5} \plotone{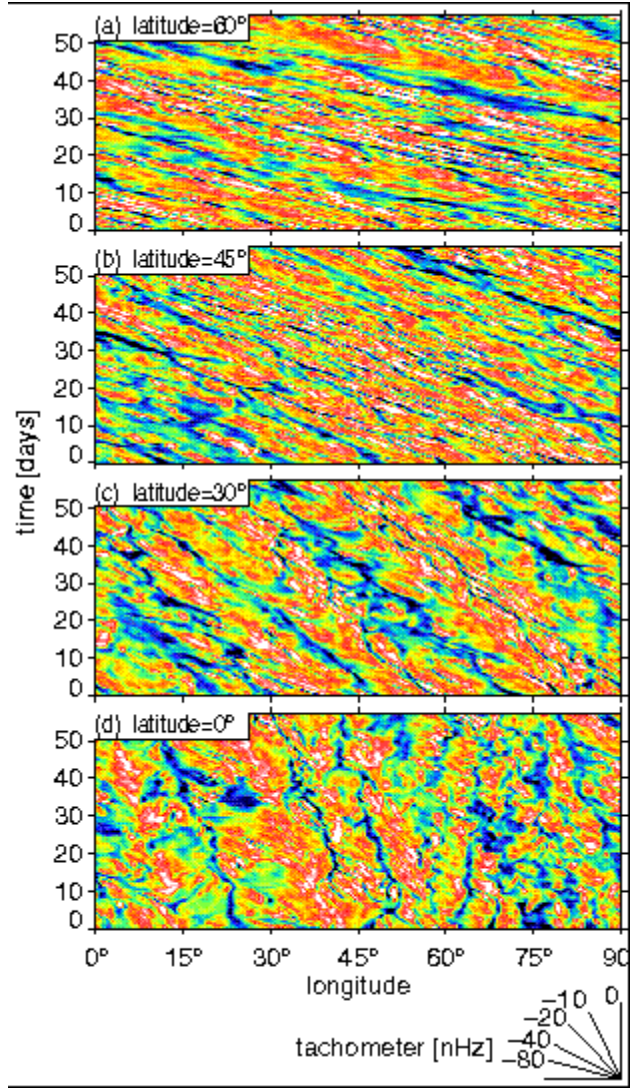}
  \caption{Near-surface radial velocity structures at four latitudes from
  Case~\stwo, plotted as a function of time and longitude.  As labeled, the
  four panels correspond to latitudes of~$0^\circ$,~$30^\circ$,~$45^\circ$,
  and~$60^\circ$.  The retrograde propagation rate of these features,
  quantified by the tachometer, is a reflection of the no-slip differentially
  rotating lower boundary.}  \label{fig:evol_shsl_xt_s2}
\end{figure}

\begin{figure*}
  \epsscale{1.0} \plotone{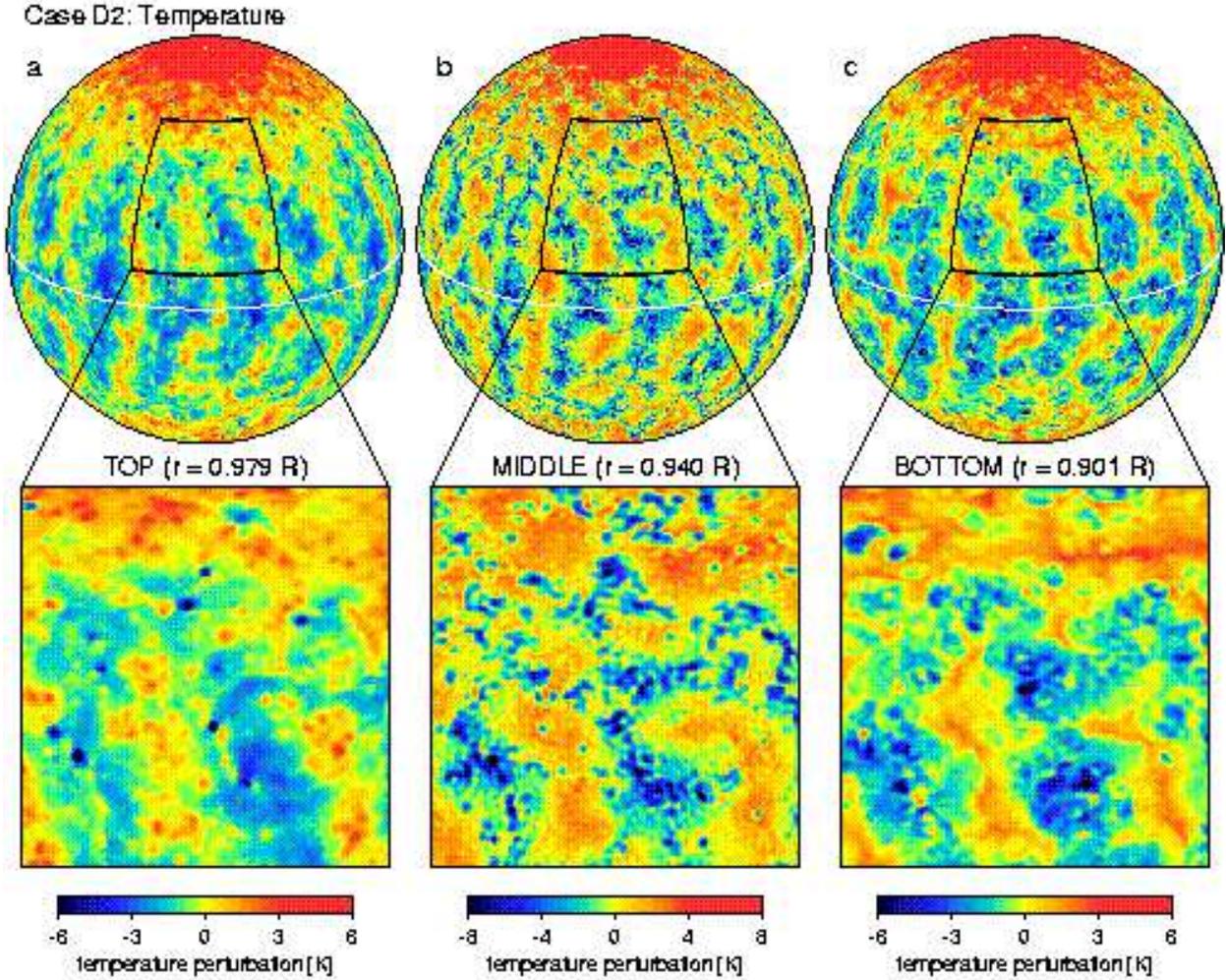}
  \notetoeditor{Figure 10 should be a full-width double-column figure.}
  \caption{Similar to Fig.~\ref{fig:shsl_d2_ur}, but showing the horizontal
  structure of the temperature perturbation sampled near the ($a$) top, ($b$)
  middle, and ($c$) bottom of the domain, with the mean temperature for each
  level removed.  Orange-red colors denote warmer fluid, green-blue colors
  denote cooler fluid.}  \label{fig:shsl_d2_t}
\end{figure*}

\begin{figure}
  \plotone{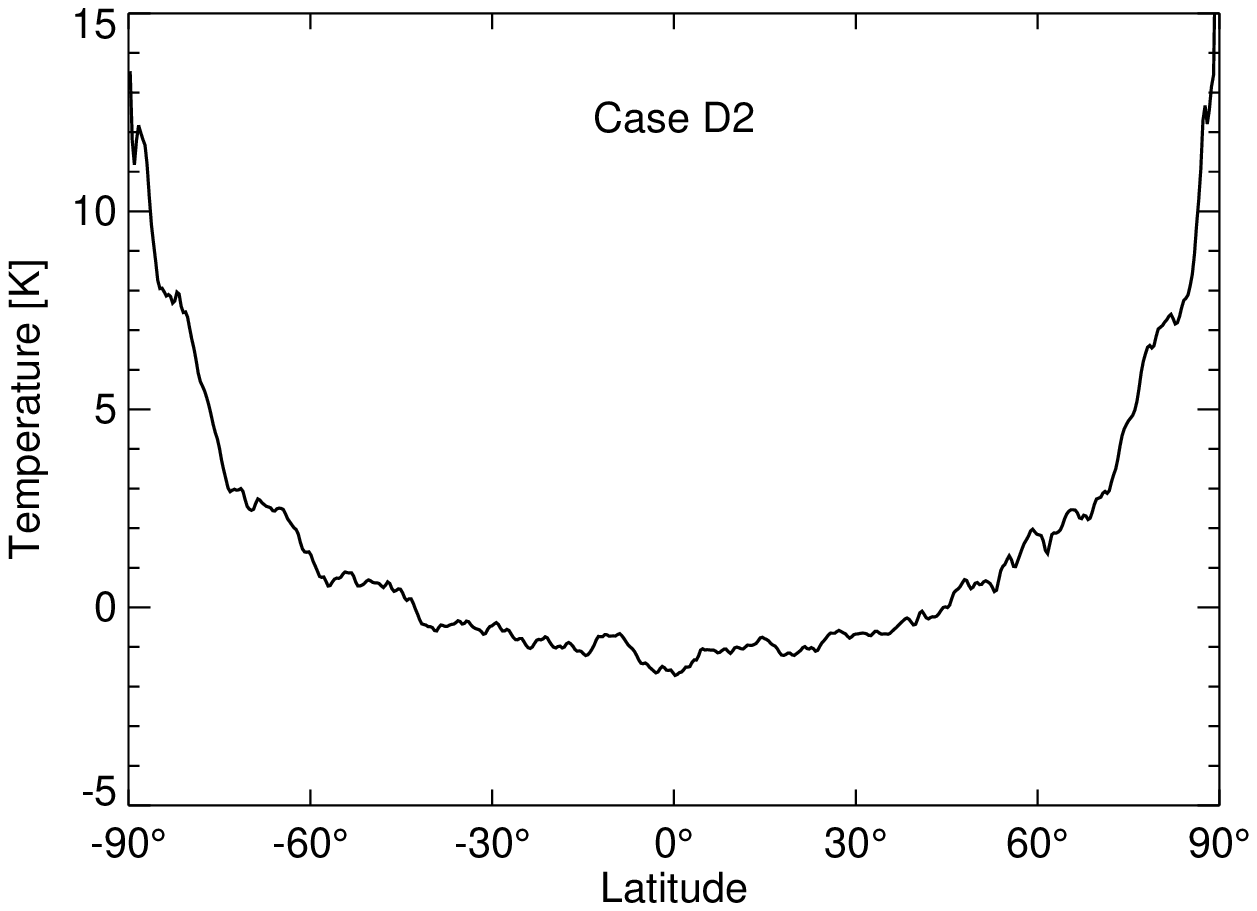}
  \caption{The zonally averaged fluctuating temperature near the top of the
  domain ($r$=0.979~$R$) for Case~\dtwo.}  \label{fig:shsl_d2_avt}
\end{figure}

The convective flow patterns in these simulations are intricate, containing
complex evolving structures occurring on multiple size scales.  We illustrate
typical velocity patterns in Figures~\ref{fig:shsl_s2_ur}
and~\ref{fig:shsl_d2_ur} by showing the radial velocity for Cases~\stwo{}
and~\dtwo{} at several depths within each domain.  The horizontal structure of
the radial velocity fields realized in Cases~\sone{} and~\dthr{} are
qualitatively similar to Cases~\stwo{} and~\dtwo{}, respectively.

Figures~\ref{fig:shsl_s2_ur}$a$ and~\ref{fig:shsl_d2_ur}$a$ show that the
largest scale of convection visible near the top of both Cases~\stwo{}
and~\dtwo{} is associated with a connected network of downflow lanes
(green-blue colors) having a spatial scale of about 200~Mm.  The large areas
enclosed by the downflow lanes each contain several smaller-scale upflows
(orange-red tones) measuring about 15--30~Mm across.  Although the upflow
cells in the shallow shell (Case~\stwo) tend to be slightly larger than those
in the deeper shell (Case~\dtwo), this general surface pattern of a network of
connected downflow lanes enclosing several distinct smaller upflows appears to
be a robust property of convection within our thin shells.  The size of the
smaller upflows in these simulations is approximately 20~Mm, similar to the
horizontal size scale of solar supergranulation.

Figures~\ref{fig:shsl_s2_ur}$b$,$c$ and~\ref{fig:shsl_d2_ur}$b$,$c$ illustrate
how the horizontal planforms change with depth within Cases~\stwo{} and~\dtwo.
As the downwelling fluid reaches deeper layers, the degree of connectivity of
the downflow network appears to decrease, since the vertical velocities along
each downflow lane becomes much less uniform.  Near the bottom of each domain,
the downflows have fragmented into more isolated and compact plume-like
structures while retaining a vestige of the connected network of downflow
lanes seen near the surface.  This narrowing of scale with depth appears to be
related to the larger densities found near the bottom of each domain.

The strongest downflow lanes in the equatorial region of Case~\dtwo{} possess
a noticeable north-south orientation that is somewhat reminiscent of the
banana-cell modes evident in more laminar spherical shell convection
simulations (e.g.~simulation LAM in \citealt{mie2000}).  However, the columnar
structures seen in Case~\dtwo{} are much thinner, have more spatial variation,
and evolve more quickly than the banana cells in the more laminar cases, as a
result of the strong small-scale convection driven throughout the domain.

The smaller-scale upflows visible near the upper boundary of these simulations
also change their character with depth.  The distinct upflows enclosed by
downflow lanes seen in the upper layers gradually become more uniform, forming
broad regions of upwelling fluid surrounded by an incomplete network of
downflow lanes.  Near the lower boundary, these broad upflow regions have
largely disappeared, with the fastest upward velocities occurring in shrouds
surrounding each of the strong downflows in response to the impenetrable lower
boundary that exists at the bottom of the domains.  Figure~\ref{fig:posl_ur}
contains vertical cuts of the radial velocity field showing the variation with
latitude and radius for Cases~\stwo{} and~\dtwo.  The radial structure of the
upflow regions is most evident in Case~\dtwo{} (Fig.~\ref{fig:posl_ur}$b$),
wherein the smaller-scale yet faster upflows are only evident near the upper
boundary of the domain.

Figure~\ref{fig:evol_shsl_s2_ur} shows the time evolution of the radial
velocity field of Case~\stwo.  Such time series illustrate the tendency for
features in the flow field to be systematically advected by velocity patterns
possessing a larger scale. One small upflow (indicated by the arrow) as well
as the downflow lane immediately to its left are both advected laterally by
the horizontal outflow motions associated with the broader cell. The center
of this broad outflow cell is indicated by a cross.  Such lateral transport
of velocity features by larger scales of convection is most apparent in movie
sequences showing the time evolution of the radial velocity field.

In addition to the network of downflow lanes possessing larger spatial scales
than are observed at the solar photosphere, we also find that these structures
persist for longer time intervals than their counterparts on the sun.  For
example, the time series shown in Figure~\ref{fig:evol_shsl_s2_ur} illustrates
that the prominent downflow network visible near the upper surface of
Case~\stwo{} evolves on time scales longer than about 10 days.  Similar
evolutionary time scales are evident in Case~\dtwo{} as well.  Such
discrepancies between the numerical simulations and the observations is not
surprising, since the level of turbulence is much lower (by several orders of
magnitude) in our simulations than for the sun.  Furthermore, at 0.98~$R$ the
upper boundaries of the simulations are located below the radius of the
photosphere, and consequently may not accurately depict the convective cell
patterns that would exist above this radius.

In Figure~\ref{fig:evol_shsl_xt_s2} we present space-time diagrams of radial
velocity sampled in time for four specific latitudes (and all longitudes) at
a radius of 0.978~$R$ for Case~\stwo, plotted with respect to the rotation
rate of the computational frame $\Omega_0$.  The advection of radial velocity
structures appear as slanted features in each panel, with the higher
latitudes rotating more slowly (retrograde) than those near the equator.  The
mean advection rate of these radial velocity patterns evident in the figure
approximately equals the differential rotation rate of the fluid for the same
depth; consequently, the latitudinal variation of angular velocity shown in
the figure results largely from the imposed differential rotation at the
lower boundary.  However, it is interesting to note from
Figure~\ref{fig:evol_shsl_xt_s2} that the pattern speed of the advected
structures does vary with longitude, as structures having the same latitude
but separated in longitude may exhibit slightly different rates or
propagation.

At the mid-point in radius of their respective domains, the rms radial
velocities are measured to be 105~m~s$^{-1}$ for Case~\stwo{} and
140~m~s$^{-1}$ for Case~\dtwo, with overturning times on the order of
6--10~days, depending on the shell depth.  These overturning times suggest
that the large-scale convective pattern may be weakly sensitive to rotational
effects, as both Cases~\stwo{} and~\dtwo{} are rotating at the solar-like
mean rate of one rotation per 28~days.  We will discuss the rotational
influence of these convective overturning motions in more detail in
\S\ref{sec:axienergy}.

Figure~\ref{fig:shsl_d2_t} shows the fluctuating temperature field for
Case~\dtwo, after removing the spherically symmetric component of the
temperature at each radius.  We find that the locations of the large regions
of warm and cool temperatures correlate well with the locations of the broad
upflows and narrow downflow lanes visible in the radial velocity field shown
in Fig.~\ref{fig:shsl_d2_ur}, with the primary difference being that the
cooler regions tend to be much broader than their associated radial velocity
counterparts.  This feature is somewhat surprising given the equal thermal and
viscous diffusivities (i.e.~$P_r=1$) within Case~\dtwo, and is likely
indicative of the degree to which the cool fluid associated with the strong
downflow lanes undergoes mixing.

Near the equator, the temperature field is dominated by columns of alternating
warm and cold fluid associated with the weak banana-cell-like structures
visible in the radial velocity images. In both temperature and radial
velocity, these structures are sheared slightly by the differential rotation
within the domain, and extend up to about $\pm30^\circ$ of latitude.  (For
comparison, the cylinder tangent to the inner radius intersects the outer
radius at about $\pm20^\circ$ of latitude.)  These large-scale columnar
temperature structures are broken up by smaller-scale variations on the
temperature field which also tend to correlate well with some of the
small-scale radial velocity features. For example, the localized hot and cold
spots particularly evident in the close-up views of
Figure~\ref{fig:shsl_d2_t}$a$ are coincident with some of the fastest fluid
motions visible in the radial velocity image of Figure~\ref{fig:shsl_d2_ur}.

Near the upper boundary of Case~\dtwo, there exists a significant latitudinal
temperature contrast between the equator and the poles, as shown in
Figure~\ref{fig:shsl_d2_avt}. The temperatures in the near-polar regions are
about 10--15~K warmer than near the equator, although over half of the total
equator-to-pole difference occurs within $10^\circ$ of the pole.  As stated
earlier, we believe that many characteristics of the fluid in the near-polar
regions are most likely artifacts of the four-fold azimuthal periodicity
imposed in these simulations, and should be interpreted with care.

\subsection{Axisymmetric Flow Patterns}
\label{sec:axiflows}

\begin{figure}
  \epsscale{0.9} \plotone{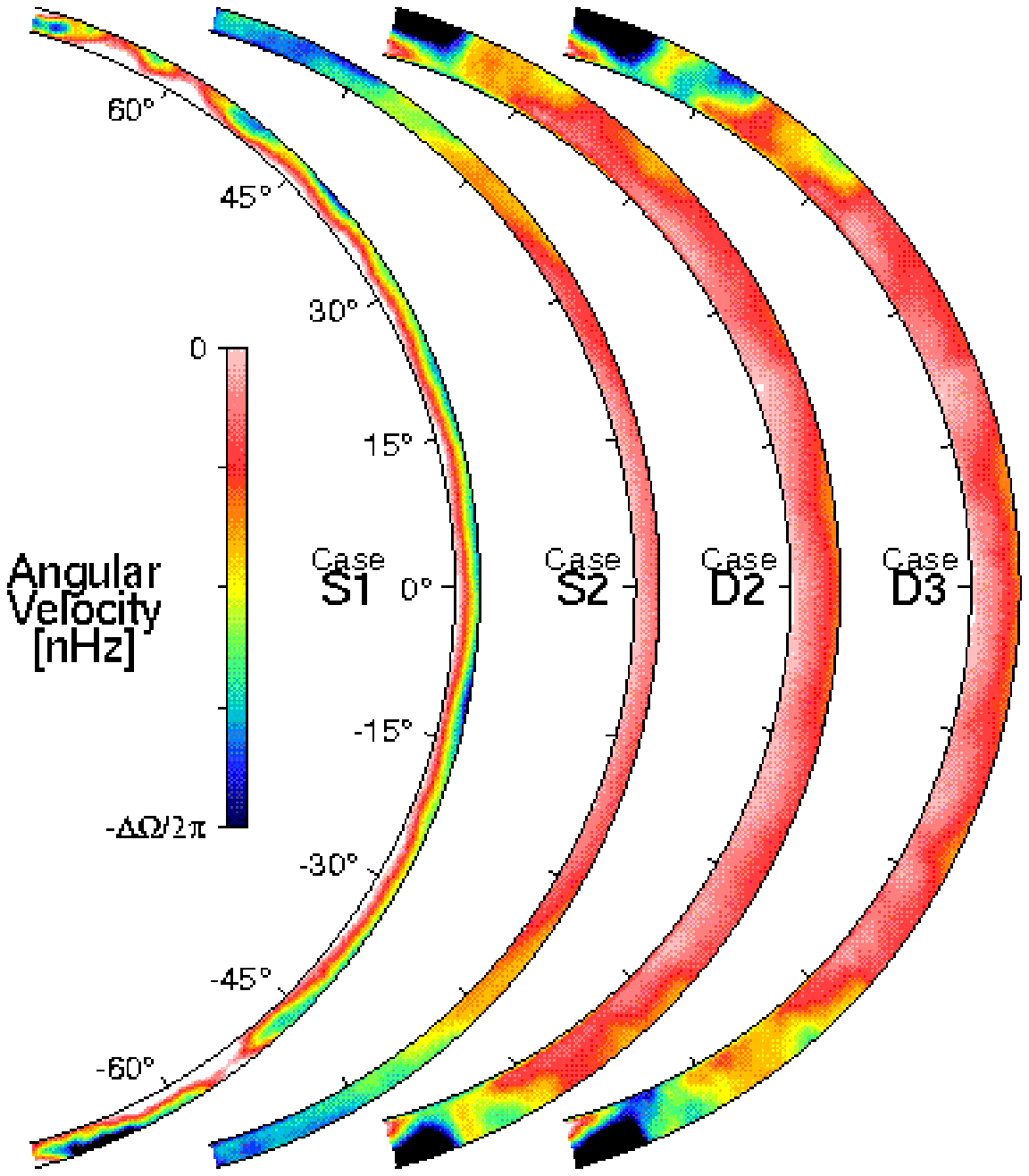}
  \caption{Angular velocity $\Omega/2\pi$ relative to the rotating coordinate
  system as a function of latitude and radius for each case, averaged over
  longitude and time.  A no-slip differentially rotating lower boundary is
  imposed in Cases~\stwo,~\dtwo, and~\dthr.  That imposed angular velocity
  decreases from 0~nHz at the equator to about $-120$~nHz at a latitude of
  $75^\circ$. Case~\sone{} has a uniformly rotating no-slip lower boundary.
  The images are scaled between 0 and $-\Delta\Omega/2\pi$, with a limit of
  $-20$~nHz for Case~\sone{} and $-120$~nHz for Cases~\stwo,~\dtwo,
  and~\dthr.}  \label{fig:azav_dr}
\end{figure}

\begin{figure}
  \epsscale{0.9} \plotone{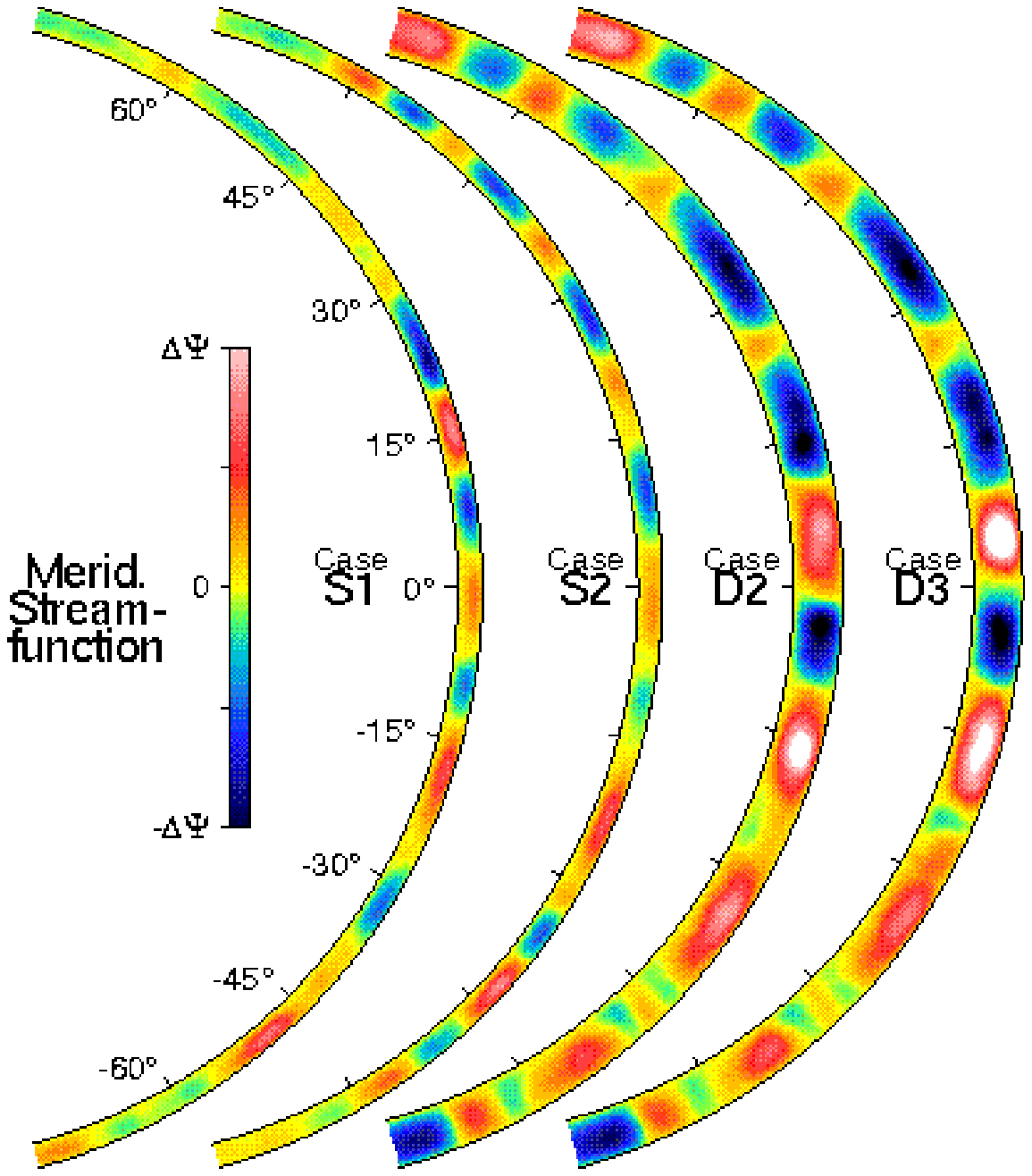} 
  \caption{Mass flux streamfunction $\Psi$ associated with the meridional
  circulation for each case, averaged over longitude and time.  Contours of
  constant $\Psi$ coincide with velocity streamlines, with positive values
  representing flow in the clockwise sense.  The limiting values
  $\pm\Delta\Psi$ are $\pm 4\times10^{23}$ cm$^3$~s$^{-1}$ for Cases~\sone{}
  and~\stwo, and $\pm 6\times10^{23}$ cm$^3$~s$^{-1}$ for Cases~\dtwo{}
  and~\dthr.  The corresponding flow velocities in the meridional plane are of
  order 50 and 75~m~s$^{-1}$ respectively.}
\label{fig:azav_str}
\end{figure}

The time-averaged axisymmetric (longitudinally averaged) profile of angular
velocity $\Omega/2\pi$ achieved within each simulation is shown in
Figure~\ref{fig:azav_dr}, with the quantity $\Omega$ now corresponding to the
angular velocity in the rotating frame after the angular velocity of the
computational frame $\Omega_0$ is subtracted out.  The no-slip rotating
boundary is imposed such that equatorial rate corresponds to the rotation rate
of the computational frame, such that $\Omega/2\pi=0$ at these points.  The
differential rotation of the lower boundary imposed in Cases~\stwo,~\dtwo,
and~\dthr{} decreases from $\Omega/2\pi=0$~nHz at the equator to about
$\Omega/2\pi=-120$~nHz at a latitude of $75^\circ$, and is similar in contrast
and functional form to the latitudinal variation of the photospheric plasma
rate measured by \citet{sno1984}.  For comparison purposes, the no-slip lower
boundary imposed in Case~\sone{} is uniformly rotating at $\Omega/2\pi=0$~nHz.

With the exception of the polar regions (not shown in the figure), the angular
velocity profile within each simulation is retrograde with respect to the
rotating coordinate system, with the fastest rotation rates occurring near the
bottom of each shell at any given latitude.  Figure~\ref{fig:azav_dr} shows
that Cases~\sone{} and \stwo{} possesses a largely constant negative radial
gradient of angular velocity with radius at each point (see also
Fig.~\ref{fig:radcuts_scases}), with the overall magnitude of $\Omega$
determined by the rotation rate imposed at the corresponding latitude on the
lower boundary.  Within Cases~\dtwo{} and \dthr, the negative radial gradients
in angular velocity throughout the bulk of each shell are smaller, except
within the thin viscous boundary layers located near the lower boundaries of
both simulations.

The regions poleward of $\pm 75^\circ$ of latitude are not shown in
Figure~\ref{fig:azav_dr} due to the high angular velocities present there.
Even though these regions exhibit reasonable (linear) zonal velocities, the
short moment arm at such high latitudes produces higher values of $\Omega$
than are seen elsewhere within the domain at lower latitudes.  Furthermore,
the polar regions of Cases~\sone, \stwo, and~\dtwo{} are dominated by effects
related to the four-fold azimuthal periodicity imposed in these simulations,
which disallows flows across the poles and limits the physical size of
convective structures.  Such effects are likely to produce dynamics which may
not be otherwise present in analogous simulations without the four-fold
azimuthal symmetry.  Although all longitudes are computed in Case~\dthr, it
was started from an evolved solution of Case~\dtwo{} and still shows some
effects of the angular periodicity imposed in that simulation.

The time-averaged meridional circulation achieved in our simulations is
illustrated in Figure~\ref{fig:azav_str}, where the profile of the meridional
velocity streamfunction $\Psi$ for each simulation is shown.  Most of the
kinetic energy associated with the meridional flows is contained in
$15^\circ$-wide latitudinal rolls distributed across the mid-latitude and
equatorial regions.  These rolls span the entire depth of the domain both in
the shallow- and deep-shell simulations and possess typical flow speeds of
50--75~m~s$^{-1}$.  Rolls having poleward velocities in the surface layers
appear to be preferred over rolls having the opposite rotational sense,
especially in the mid- and low-latitude regions, as they are generally more
extended horizontally and possess faster fluid velocities.

Alternatively, one could view these meridional flow patterns as an array of
symmetric clockwise and counterclockwise rolls superimposed on a single
meridional cell in each hemisphere.  We shall see that the magnitude of the
single hemispheric meridional cell is larger relative to the smaller-scale
rolls when the lower boundary is differentially rotating, as in Cases~\stwo,
\dtwo, and~\dthr.  We note, however, that because these simulations have
impenetrable boundaries, any meridional flow achieved in our simulations is
forced to close within the (relatively shallow) domain.  In this regard, the
profiles of meridional circulation obtained in these four simulations are
likely to be unrealistic, as the sun obviously does not contain impenetrable
boundaries above and below the near-surface shear layer.

\subsection{Angular Momentum Balance}
\label{sec:axienergy}

In order to understand the maintenance of the differential rotation profiles
shown in Figure~\ref{fig:azav_dr}, we will now examine the angular momentum
balance within each of the four simulations in more detail.  We will consider
only axisymmetric quantities, defined by
\begin{equation}
  \widehat{A}(r,\theta) = \frac{1}{2\pi} \int_0^{2\pi} d\phi \, A(r,\theta,\phi),
\end{equation}
where the hat $\widehat{\;}$ signifies that the quantity $A(r,\theta,\phi)$
has been averaged over longitude~$\phi$ and further over time.  Consequently,
this operation allows the decomposition
\begin{equation}
  A(r,\theta,\phi) = \widehat{A}(r,\theta) + A'(r,\theta,\phi) \qquad\text{such
  that}\qquad \widehat{A'}=0,
\end{equation}
where the prime on $A'$ denotes the non-axisymmetric component of $A$.

By multiplying the zonal component of the momentum evolution equation
(\ref{eq:evol1}) by $r\sin\theta$, we can derive an equation describing the
evolution of the angular momentum $\widehat{L}=\bar{\rho} r \sin\theta\,
\widehat{u_\phi}$, written in conservative form as
\begin{equation}
  \frac{\pd \widehat{L}}{\pd t} = -\deldot \bs F,
\end{equation}
where $\bs F$ is the angular momentum flux vector.  In a statistically steady
state, we must have $\partial \widehat{L} / \partial t=0$, such that $\deldot
\bs F=0$ throughout the domain.  Symbolically, the components of the total flux
in the radial and latitudinal directions can be written as
\begin{align}
  F_r &= F_r^{\text{DIF}} + F_r^{\text{RS}} + F_r^{\text{MC}} \label{eq:fr} \\
\intertext{and}
  F_\theta &= F_\theta^{\text{DIF}} + F_\theta^{\text{RS}} +
  F_\theta^{\text{MC}}, \label{eq:ftheta}
\end{align}
where we have used the abbreviations in capital letters to signify the
contributions to the total angular momentum flux resulting from viscous
diffusion (DIF), non-axisymmetric Reynolds stresses (RS), and axisymmetric
meridional circulation (MC):
\begin{align}
  F_r^{\text{DIF}} &= - \bar{\rho} r \sin\theta \left[ \nu r \frac{\pd}{\pd r}
    \left( \frac{\widehat{u_\phi}}{r} \right) \right] \\
  F_r^{\text{RS}} &= \bar{\rho} r \sin\theta\, \widehat{u'_r\, u'_\phi}
    \label{eq:frrs} \\
  F_r^{\text{MC}} &= \bar{\rho} r \sin\theta\, \widehat{u_r} (
    \widehat{u_\phi} + \Omega_0 r \sin\theta ) \\
\intertext{and}
  F_\theta^{\text{DIF}} &= - \bar{\rho} r \sin\theta \left[ \nu
    \frac{\sin\theta}{r} \frac{\pd}{\pd\theta} \left(
    \frac{\widehat{u_\phi}}{\sin\theta} \right) \right] \\
  F_\theta^{\text{RS}} &= \bar{\rho} r \sin\theta\, \widehat{u'_\theta\,
    u'_\phi} \\
  F_\theta^{\text{MC}} &= \bar{\rho} r \sin\theta\, \widehat{u_\theta} (
    \widehat{u_\phi} + \Omega_0 r \sin\theta ).
\end{align}

By examining the contributions to the total angular momentum flux from each of
these components, we will show that the radial gradients of angular velocity
realized in each of the four simulations are supported against diffusion by
Reynolds stresses associated with non-axisymmetric convective motions.  This
behavior can be thought of as the tendency of convective fluid elements to
partially conserve their angular momentum per unit mass $\lambda = \Omega r^2
\sin^2\theta$ as they move toward or away from the axis of rotation.  As
suggested by \citet{fou1975}, constancy of $\lambda$ along radial lines may
also explain why surface magnetic tracers on the sun have a faster rotation
rate relative to the surface fluid, if one assumes that the magnetic features
are anchored at a radius slightly below the photosphere where the rotation
rate is faster.  \citet{gil1979} tested this notion by numerically modeling
Boussinesq convection confined to a thin shell, and found that angular
momentum was roughly conserved along local radii for the case of an
incompressible fluid.  They demonstrated that the convective motions were able
to transport angular momentum inward, thereby maintaining the negative radial
gradient of rotation rate with radius.  The models presented here indicate
that compressible convection behaves similarly.

\begin{figure*}
  \plotone{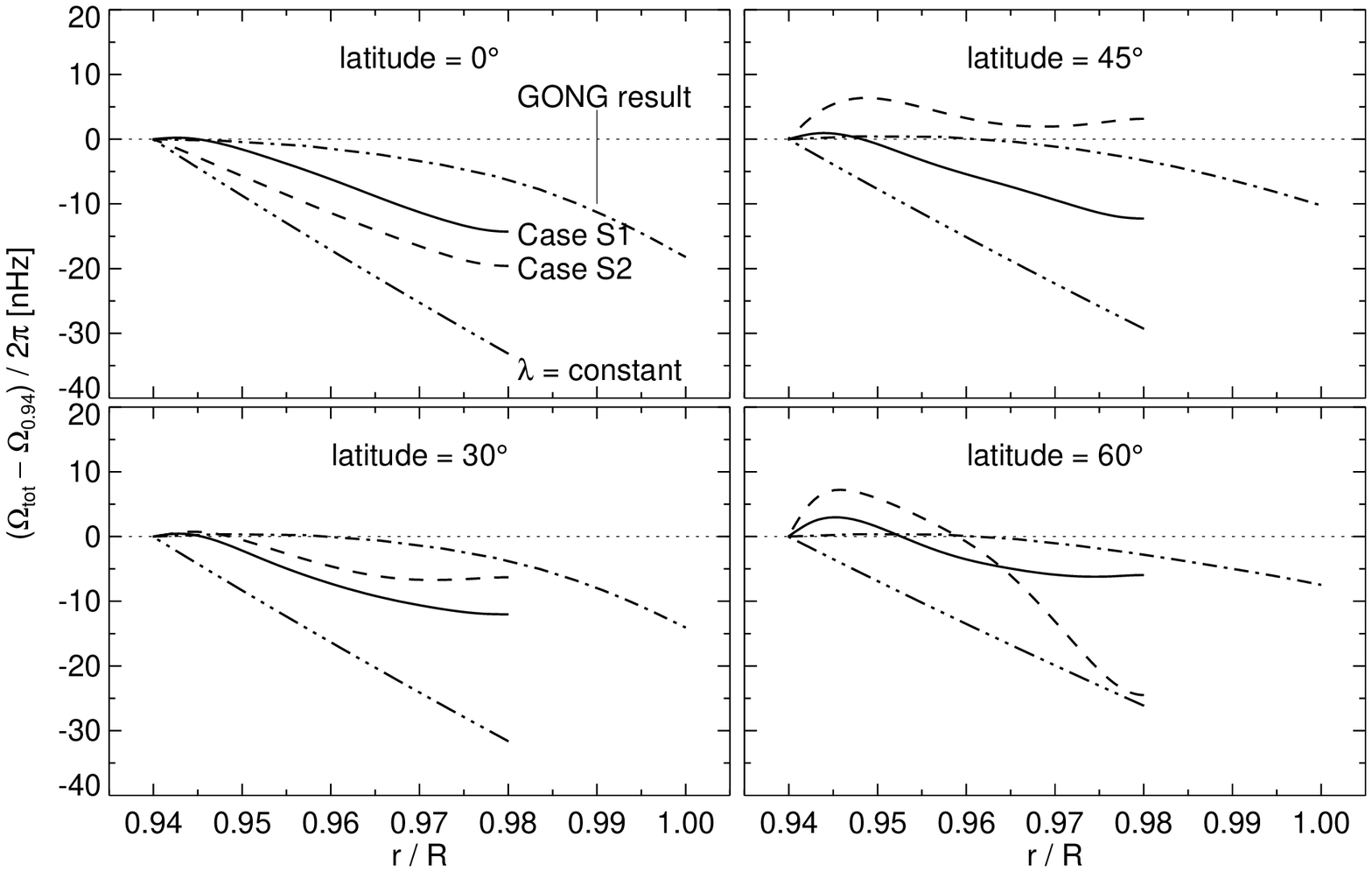}
  \notetoeditor{Figure 14 should be a full-width double-column figure.}
  \caption{Angular velocity profiles for Cases~\sone{} ({\sl solid})
  and~\stwo{} ({\sl dashed}) as a function of radius for latitudes
  $0^\circ$,~$30^\circ$,~$45^\circ$, and~$60^\circ$ as indicated.  The
  dash-dot-dot-dot line in each panel represents the angular velocity of a
  radially moving fluid parcel for which its angular velocity per unit mass
  $\lambda$ is conserved, while the dash-dot line corresponds to the GONG data
  plotted in Fig.~\ref{fig:gong}.}  \label{fig:radcuts_scases}
\end{figure*}

\begin{figure*}
  \plotone{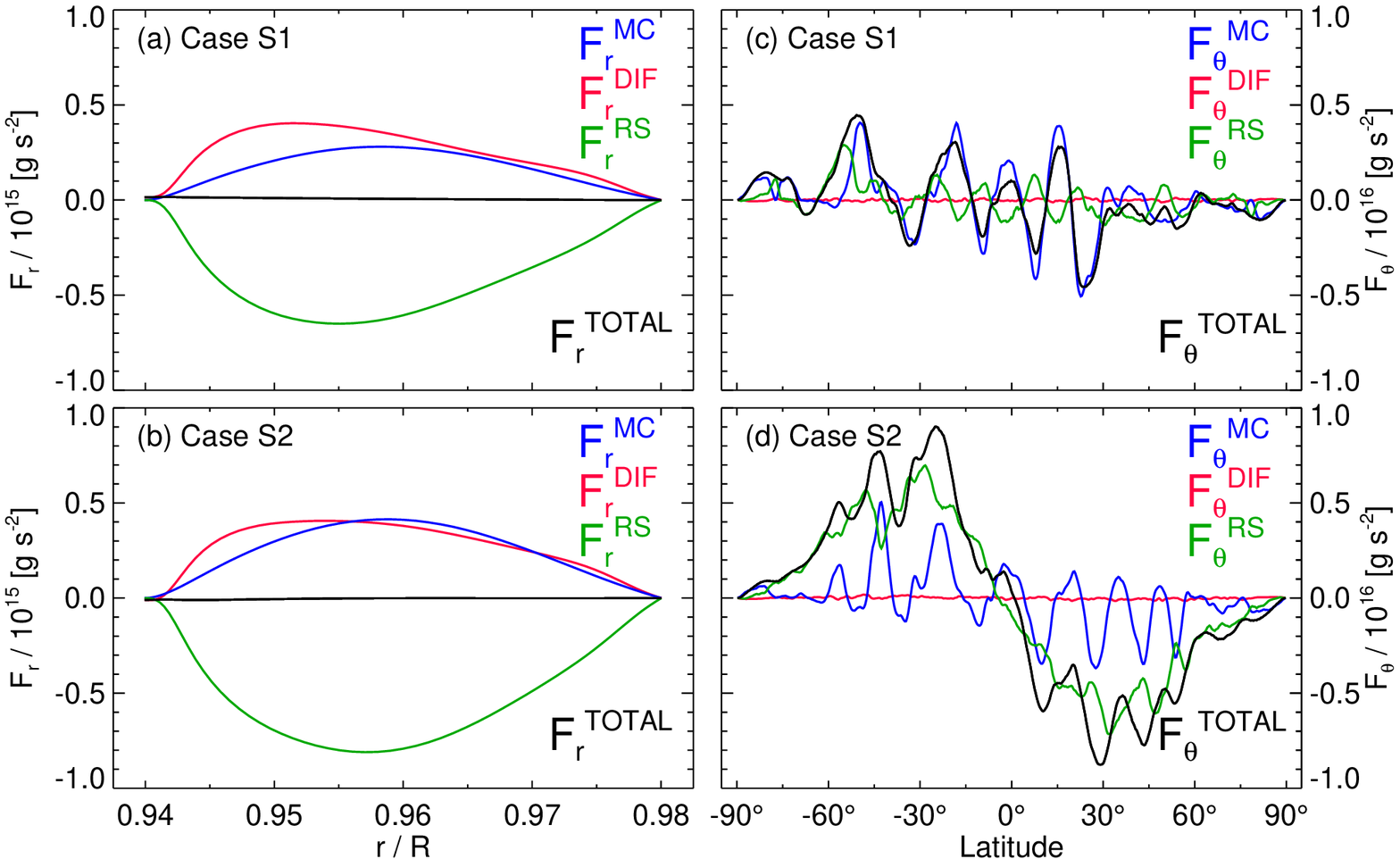}
  \notetoeditor{Figure 15 should be a full-width double-column figure.}
  \caption{The components of the radial ($a$) and ($b$) and latitudinal ($c$)
  and ($d$) angular momentum fluxes for Cases~\sone{} and~\stwo.  Positive
  values of $F_r$ and $F_\theta$ indicate upward and southward transports,
  respectively.}  \label{fig:amflux_s}
\end{figure*}

We begin by examining the angular momentum balance within the two
shallow-shell simulations, Cases~\sone{} and~\stwo.  These two simulations are
identical, except that Case~\stwo{} has a differentially rotating lower
boundary, whereas in Case~\sone{} it is forced to rotate uniformly.  The upper
boundary in both cases is stress-free.  From Figure~\ref{fig:azav_dr} it is
apparent that the radial angular velocity gradient $\partial\Omega / \partial
r$ is negative over a broad range of latitudes.  This effect is further
illustrated in Figure~\ref{fig:radcuts_scases}, where the angular velocity
profiles shown in Figure~\ref{fig:azav_dr} for Cases~\sone{} and~\stwo{} are
plotted as a function of radius for selected latitudes.  The dash-dot-dot-dot
lines in each panel indicate the angular velocity of an isolated fluid parcel,
assuming it were moving in a purely radial direction and conserving $\lambda$
(making $\Omega \propto r^{-2}$).

Figure~\ref{fig:amflux_s} illustrates the contributions to the total radial
and latitudinal angular momentum fluxes from each of the component fluxes
listed in equations~(\ref{eq:fr}) and~(\ref{eq:ftheta}).  The radial fluxes
of Cases~\sone{} and~\stwo{} are shown in Figures~\ref{fig:amflux_s}$a$,$b$
after integrating over latitude.  In both cases, the average angular momentum
fluxes associated with viscous diffusion $F_r^{\text{DIF}}$ and meridional
circulation $F_r^{\text{MC}}$ are positive at all radii, signifying a
transport of angular momentum toward the top of the domain.  We find that
this upward transport is balanced almost entirely by the downward transport
achieved by non-axisymmetric Reynolds stresses associated with the
nonaxisymmetric convective motions.  For this effect to occur in this manner,
the velocity correlation $\widehat{u'_r \, u'_\phi}$ must be negative on
average by equation~(\ref{eq:frrs}) for $F_r^{\text{RS}}<0$, indicating that
small-scale non-axisymmetric radial motions, such as the fast downflows, tend
to possess a retrograde tilt in the $r\phi$-plane.  Such a tendency is
exactly what one would expect if non-axisymmetric convective motions were to
partially conserve their angular momentum per unit mass $\lambda$ in radial
motion.

In the sun, while $\partial\Omega / \partial r$ is negative between
0.94--0.98~$R$, its magnitude is much smaller than in Cases~\sone{} and~\stwo,
as shown in Figure~\ref{fig:radcuts_scases}.  In the figure, the dash-dot
lines denote the run of $\Omega$ with radius as determined from
helioseismology.  Since the viscosity of the sun is much lower, and since the
solar meridional circulation is of the same magnitude as in our shallow-shell
cases, the tendency for non-axisymmetric convective motions to conserve
$\lambda$ is likely to be less effective in the sun between 0.94--0.98~$R$
than in the numerical simulations presented here. The helioseismic results
surfaceward of 0.98~$R$, also plotted in the figure, suggest that the
smaller-scale convective motions located closer to the photosphere may tend to
conserve their angular momentum more fully, thereby maintaining a steeper
angular velocity gradient $\partial\Omega / \partial r$ in this region.

Figure~\ref{fig:radcuts_scases} shows that both Cases~\sone{} and~\stwo{}
possess viscous boundary layers near the bottom of each domain, especially at
higher latitudes.  These Ekman-type layers are formed in response to the zonal
velocity imposed at the lower boundary in each case and have no physical
analog in the sun.  Within this boundary, the influence of viscous dissipation
is large enough to flatten out the angular velocity gradient at low latitudes
and even produce a positive $\partial \Omega/ \partial r$ at the higher
latitudes (e.g.~at $60^\circ$ in~Fig.~\ref{fig:radcuts_scases}).  We shall see
that this effect is even more pronounced in Cases~\dtwo{} and~\dthr.

The latitudinal flux of angular momentum in Cases~\sone{} and~\stwo{} is
shown in Figures~\ref{fig:amflux_s}$c$,$d$ after integrating in radius.  We
find that Case~\stwo, having a differentially rotating lower boundary, has a
much stronger poleward transport of angular momentum relative to Case~\sone{}
in which the lower boundary is uniformly rotating.  This effect is an
expected consequence due the additional angular momentum being removed from
the system at the lower boundary in the near-polar regions of Case~\stwo,
since the lower boundary in this case is differentially rotating rather than
uniformly rotating as in Case~\sone.  Such an enhanced poleward transport of
angular momentum within Case~\stwo{} is achieved primarily by the convective
motions, and to a lesser degree by the meridional circulation, as illustrated
by the magnitudes of the fluxes $F_\theta^{\text{RS}}$
and~$F_\theta^{\text{MC}}$ in Figure~\ref{fig:amflux_s}$d$.  A recent
analysis of photospheric velocity images suggests that there may be
observational evidence of a similar effect occurring on the sun, where the
Reynolds stresses associated with solar supergranulation yield a poleward
transport of angular momentum within the near-surface layers of the sun
\citep{hat2001}.

Superimposed on the broad poleward transport, the angular momentum fluxes
$F_\theta^{\text{RS}}$ and~$F_\theta^{\text{MC}}$ possess significant
latitudinal fluctuations that are directly correlated with the axisymmetric
rolls of Figure~\ref{fig:azav_str}.  We find that the contributions to
$F_\theta^{\text{MC}}$ from the latitudinal rolls is partially offset by
matching contributions to $F_\theta^{\text{RS}}$, such that the net angular
momentum transport by the rolls is somewhat limited.

\begin{figure*}
  \plotone{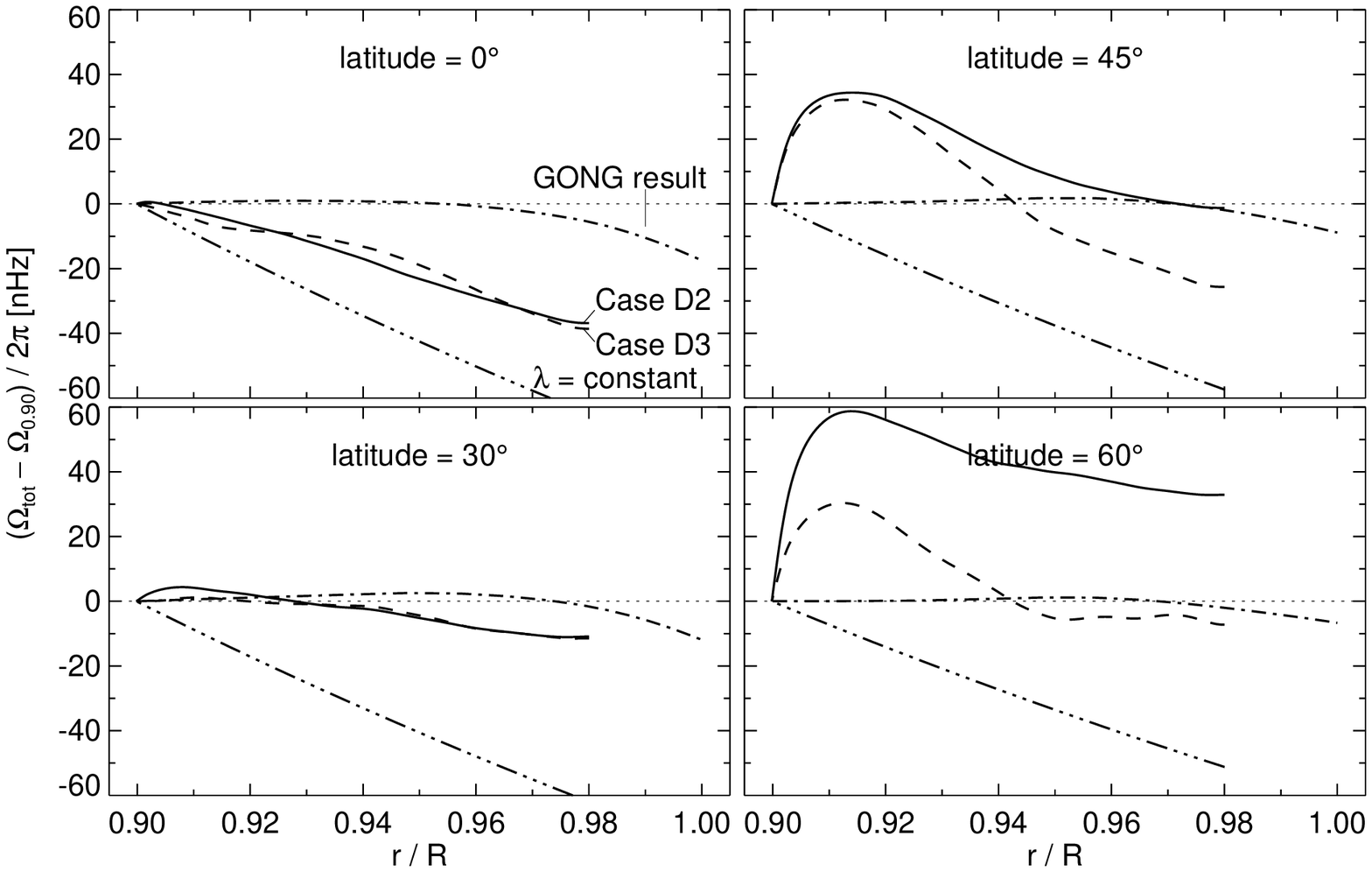}
  \notetoeditor{Figure 16 should be a full-width double-column figure.}
  \caption{Similar to Fig.~\ref{fig:radcuts_scases}, except for the deep-shell
  simulations, Cases~\dtwo{} ({\sl solid}) and~\dthr{} ({\sl dashed}).  The
  dash-dot-dot-dot line in each panel represents the angular velocity of a
  radially moving fluid parcel for which its angular velocity per unit mass
  $\lambda$ is conserved, while the dash-dot line corresponds to the GONG data
  plotted in Fig.~\ref{fig:gong}.}  \label{fig:radcuts_dcases}
\end{figure*}

\begin{figure*}
  \plotone{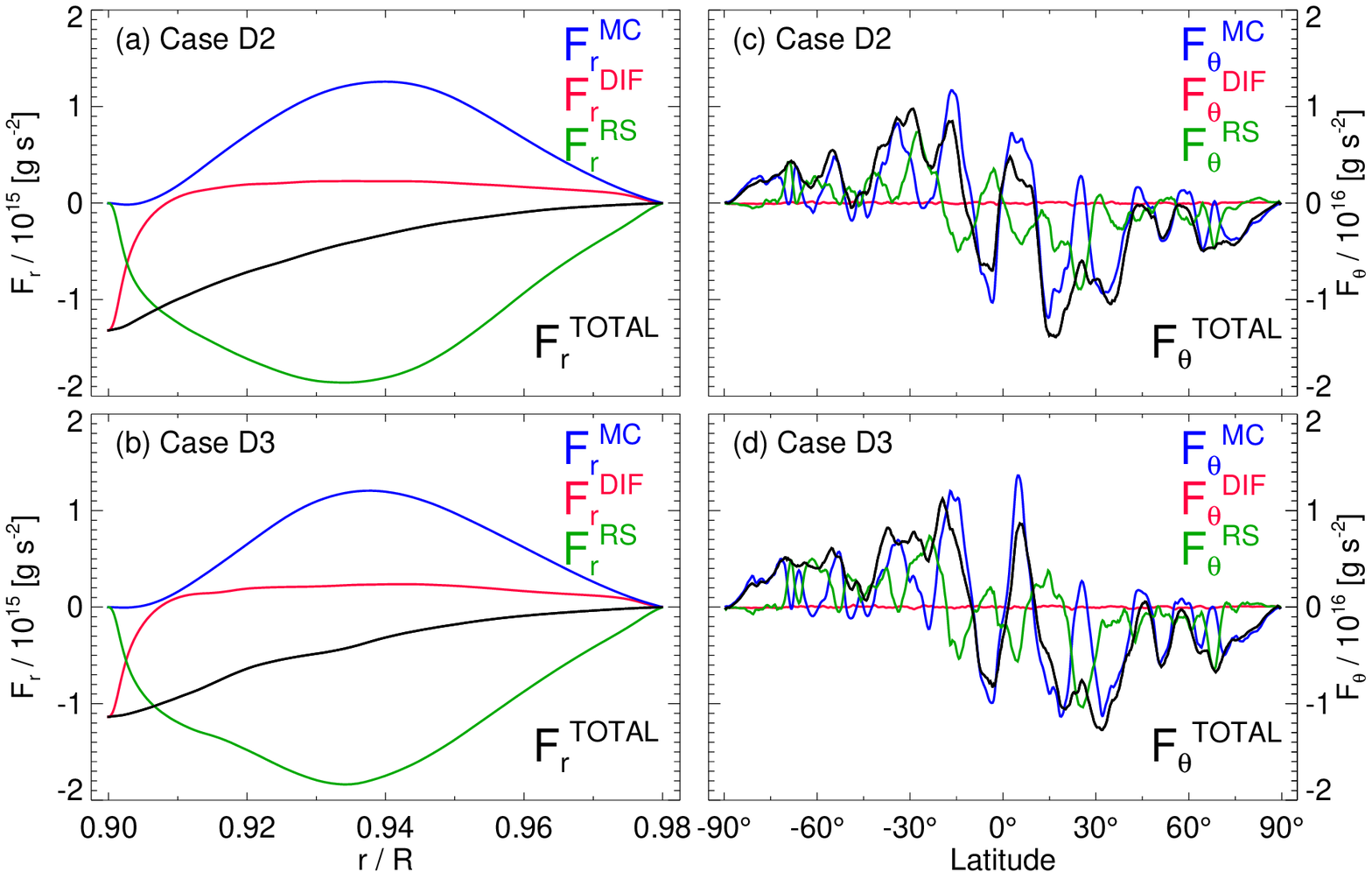}
  \notetoeditor{Figure 17 should be a full-width double-column figure.}
  \caption{The components of the radial ($a$) and ($b$) and latitudinal ($c$)
  and ($d$) angular momentum fluxes for Cases~\dtwo{} and~\dthr.  Positive
  values of $F_r$ and $F_\theta$ indicate upward and southward transports,
  respectively.}  \label{fig:amflux_d}
\end{figure*}

Turning to the deep-shell simulations of Cases~\dtwo{} and~\dthr, we find that
the angular velocity $\Omega$ possesses radial gradients similar to those
achieved within Cases~\sone{} and~\stwo.  Radial cuts through each domain,
such as those presented in Figure~\ref{fig:radcuts_dcases}, show that the
radial gradient of angular velocity is negative above 0.91~$R$ for each of the
latitudes shown.  The primary difference between the shallow and deeper shells
is the magnitude of the viscous boundary layer adjacent to the lower boundary.
The radial gradients of $\Omega$ are larger in magnitude across the viscous
boundary layer for the deeper shells, causing the diffusive transport of
angular momentum down the radial gradient of $\Omega$ to be correspondingly
greater in Cases~\dtwo{} and~\dthr{} than for Cases~\sone{} and~\stwo.  This
enhanced diffusive transport is illustrated in the curves of
$F_r^{\text{DIF}}$ of Figure~\ref{fig:amflux_d}$a$,$b$, where the greater
torques applied to the fluid via the no-slip lower boundary are evident in the
enhanced diffusive fluxes $F_r^{\text{DIF}}$ within Cases~\dtwo{} and~\dthr.
Elsewhere in the domain, viscous effects play a reduced role compared to the
radial angular momentum transport achieved by the meridional circulation and
Reynolds stresses.  As in Cases~\sone{} and~\stwo, the radial Reynolds stress
flux $F_r^{\text{RS}}$ is negative and the radial meridional circulation flux
$F_r^{\text{MC}}$ is positive on average throughout the entire domain.

Figures~\ref{fig:amflux_d}$c$,$d$ show the latitudinal fluxes of angular
momentum within Cases~\dtwo{} and~\dthr, after integrating over radius.  These
two cases indicate that angular momentum is primarily transported poleward
throughout the domain by the meridional circulation (rather than by the
Reynolds stresses as in Cases~\sone{} and~\stwo).  However, the $15^\circ$
rolls now contain more kinetic energy than their counterparts in the
shallow-shell simulations, and consequently the variations superimposed on
$F_\theta^{\text{MC}}$ are larger.  As with the shallower cases, the Reynolds
stress flux $F_\theta^{\text{RS}}$ due to these rolls tends to partially
offset the latitudinal transport of angular momentum by meridional flows.  By
comparing each of the four simulations, we suspect that the strength of the
meridional rolls in our thin-shell domains is affected by the shell depth,
which effectively limits the radial extent of any meridional circulation
within the domain.

In summary, the time scales of the largest overturning motions in our
simulations suggest that they are at least weakly influenced by rotational
effects, which in turn may enable Reynolds stresses to facilitate transport
angular momentum inward.  This inward angular momentum transport balances the
outward transport by diffusion and meridional flows on average, thereby
maintaining a negative angular velocity gradient throughout much of the layer
at low and mid latitudes.  This effect may contribute to the observed decrease
of $\Omega$ with radius in the near-surface shear layer of the sun as deduced
from helioseismic analyses.  Behavior at high latitudes is somewhat more
complex due to the presence of a viscous boundary layer near the lower
boundary, and likewise there is some uncertainty in the helioseismic
inferences about the radial gradient in $\Omega$ achieved at latitudes of
60$^\circ$ or greater.

\section{Conclusions}  \label{sec:conc}

We have presented results of three-dimensional numerical simulations of
turbulence confined to thin rotating spherical shells, seeking to understand
some of the dynamical effects that supergranular scales of motion within thin
shearing layers might have within the analogous layer located near the top of
the solar convection zone.  We have focused our analysis on the physical
processes that enable the transport of angular momentum within the thin shell
domains, in order to investigate analogous angular momentum transport on the
sun, and to determine the cause of the negative radial angular velocity
gradients shown to exist within the near-surface shear layer of the sun.

The high-resolution simulations presented here allow horizontal structures of
order 10~Mm to be explicitly resolved, thereby allowing us for the first time
to incorporate dynamical scales on the order of solar supergranulation within
global simulations of solar convection.  We find that the broad spectrum of
scales of motion, while typically much smaller in size than the largest
characteristic length scales of the convection zone, are able to influence
the large-scale dynamics of the system through their ability to transport
angular momentum within the shells on a global scale.

The vigorous convection realized in each of the four simulations presented
here is driven by imposing the solar heat flux at the lower boundary of each
domain.  Shell thicknesses of 4\% and of 8\% of the solar radius $R$ are
considered, with the upper boundary in each case located at a radius of
0.98~$R$.  We have imposed a differentially rotating no-slip lower boundary in
three of the four simulations.  Both the lower and upper boundaries are
impenetrable, and the upper boundary is forced to remain at a constant
entropy.

We find in all cases that near the middle of the domain the convection takes
the form of a connected network of fast yet narrow downflow lanes that enclose
broad regions of warmer, more slowly rising fluid.  The cells enclosed by the
downflow network typically measure 100--200~Mm across, with the lanes
themselves about 20~Mm wide.  In the deeper layers where the density is
greater, this network loses much of its horizontal connectivity, instead
forming more plume-like structures that approach the bottom of the domain, at
which point the impenetrability of the lower boundary forces the fluid to be
diverted horizontally.

Closer to the surface, the broad cells of upwelling fluid are found to segment
into several smaller upflows having a horizontal scale comparable to that of
solar supergranulation, of order 20--40~Mm.  These small-scale upflow cells
appear in both the shallow- and deep-shell simulations, suggesting that the
more superadiabatic stratification present near the top of each domain, rather
than the depth of the shell, is the primary factor that determines the
morphology of the convection near the surface.  Time series of the evolving
near-surface flow field show that both the smaller upflow cells as well as the
narrow downflow lanes are horizontally advected in a sustained fashion as they
respond to larger-scale sweeping flows that develop nearby.

Averaging the flow fields in longitude reveal that the angular velocity
decreases with radius in the low- and mid-latitude regions of each domain,
with the exception of a thin viscous boundary layer that forms near the
no-slip lower boundary in each case.  An analysis of the angular momentum
balance shows that such negative radial gradients of angular velocity are
maintained by an inward transport of angular momentum, achieved by Reynolds
stresses associated with the convective motions that balance the outward
transport of angular momentum from viscous diffusion and the global meridional
circulation.  Such an inward transport is achieved if radially moving fluid
motions, such as the broad upflows and strong downflows seen here, have the
tendency to conserve their angular momentum per unit mass while moving
radially throughout the shell.

The longitudinally averaged meridional velocity patterns take the form of a
series of $15^\circ$ latitudinal rolls that span the full radial thickness of
each shell, with cells having a poleward surface flow tending to have a
broader latitudinal extent than cells with the opposite sense.  The net
effect of these rolls is to transport angular momentum poleward, as required
by latitudinally varying angular momentum flux imposed by the differentially
rotating lower boundary.  However, we reemphasize that the profiles of
meridional circulation within these thin-shell simulations are significantly
influenced by the impenetrable radial boundaries of our simulations,
effectively forcing any circulation in the meridional plane to be completely
enclosed within the domain.  As a result, the meridional flow profiles
realized here are not expected to resemble the solar case.

The continual advance of supercomputing technology will allow simulations of
convection within thin spherical shells to be extended to deeper layers in
the future. Such models would preclude having to add angular momentum to the
system via no-slip boundary conditions, as the differential rotation and
near-surface shear layer could then be computed in a self-consistent manner.
In addition, large-scale flows driven by the small-scale convective patterns
that would not normally be confined to the near-surface layers (such as the
meridional circulation) would then be allowed to feed back on the deeper
layers below.  Preliminary attempts to construct such global models,
encompassing the bulk of the convection zone as well as a more highly
resolved layer where convection on supergranular scales can exist, are
currently underway.

We also believe the treatment of sub-grid scale (SGS) convective motions not
explicitly resolved in our simulations deserves considerable attention in the
future.  The current prescription, whereby the diffusivities are enhanced
over their thermal and molecular values, is adopted only for simplicity, and
likely does not capture all of the relevant effects of the unresolved scales
on the global dynamics.  Other treatments, such as those discussed in
\S\ref{sec:ashcode} whose functional forms depend on the shearing properties
of the resolved flows, may be more appropriate.

Nevertheless, the thin-shell simulations presented here contain highly
evolving, multi-scale convective motions that are able to efficiently
redistribute angular momentum in both radius and latitude.  Such motions are
found to maintain the radial shear within the domains, even in this idealized
environment that only approximates the near-surface shear layer of sun by
decoupling it from the bulk of the convection zone.  We also speculate that
convection in the near-surface layers of the sun may behave in a similar
fashion, maintaining the negative radial gradients in the near-surface shear
layer of the sun as deduced from helioseismology.  While we are admittedly
still far removed from directly modeling a convective layer with realistic
solar parameters, it is encouraging that the flow patterns realized in these
simulations exhibit the multiple scales of supergranulation and the more
global convection cells being deduced from local helioseismic probing of the
near-surface shear layer.

\acknowledgements

We thank Mark Miesch, Sacha Brun, and Nicholas Brummell for several rounds of
useful discussions and constructive feedback during the analysis and writing
phases of this paper.  This work was partly supported by NASA through grants
NAG~5-7996, NAG~5-8133, and NAG~5-10483, and by NSF through grant ATM~9731676.
Various phases of the simulations with ASH were carried out through NSF PACI
support of the NRAC allocation MCA93S005S at the National Center for
Supercomputing Applications (NCSA), the San Diego Supercomputing Center
(SDSC), and the Pittsburgh Supercomputing Center (PSC).  Much of the analysis
of the extensive data sets was carried out in the Laboratory for Computational
Dynamics (LCD) within JILA.

\end{document}